\documentclass[11pt,twoside]{article}
\usepackage[utf8]{inputenc}
\usepackage{graphicx,siunitx,longtable,paralist}
\usepackage{amsmath,amssymb,amsfonts,amsthm,bm}
\usepackage{subfigure}
\usepackage{subfigmat}
\usepackage{float}
\usepackage{epstopdf} 
\usepackage{multirow}
\usepackage{authblk}
\usepackage{comment}
\usepackage{appendix}
\usepackage{siunitx}
\usepackage{ctable}
\usepackage{color,soul}
\usepackage{hyperref}
\usepackage{ulem}
\usepackage[margin=2cm,font=footnotesize]{caption}

\usepackage[top=0.9 in,bottom=1.2in, left=0.8 in, right=0.8 in]{geometry}

\usepackage{titlesec}

\titleformat{\paragraph}
{\normalfont\normalsize\bfseries}{\theparagraph}{1em}{}
\titlespacing*{\paragraph}
{0pt}{3.25ex plus 1ex minus .2ex}{1.5ex plus .2ex}

\title{A study of an air-breathing electrodeless plasma thruster discharge}

\author{J. Zhou$^{1,2*}$, F. Taccogna$^2$, P. Fajardo$^1$ and E. Ahedo$^1$}

\affil[]{
$^{1}$Department of Aerospace Engineering, Universidad Carlos III de Madrid, Legan\'es, Spain
$^{2}$Istituto per la Scienza e Tecnologia dei Plasmi, Consiglio Nazionale delle Ricerche, Bari, Italy
}

\affil[]{$^*$Email: jzhou@pa.uc3m.es}
\date{}

\begin{document}
\maketitle

\begin{abstract}
Plasma chemistry of main air components is implemented in a hybrid 2D axisymmetric simulation code
to assess the air-breathing concept in an electrodeless plasma thruster. Relevant electron-heavy species collisions for diatomic molecules are included: rotational and vibrational excitation, dissociation and dissociative ionization. Plasma-wall interaction giving rise to 
associative recombination of atomic species into molecular species is included too.
As reference, the plasma thruster is operated with Xe, at a power of 300W and a mass flow of 1mg/s.
Simulations are run by injecting 1mg/s of N$_2$ and O independently for powers between 100 and 3000W. The performances and trends of  plasma response for these propellants are similar to Xe, but displaced to  powers between 1250 and 2000W. At optimum power, the thrust efficiency for
N$_2$ and O surpasses that of Xe, due to the excess of re-ionization for Xe.
Performances of 50/50 mixtures of N$_2$/O, which are a realistic composition in the ionosphere, 
are found to be linear combinations of the performances of each propellant.
Performances using O$_2$, which could be generated from associative recombination of O at the intake, are very similar to those of the atomic oxygen.
\end{abstract}


\section{Introduction}

Xenon is the most used propellant for in-space plasma electric propulsion (EP), since it offers a good trade-off of properties.
First, Xe has a low energy threshold and a high cross section for ionization, i.e. ionization requires a small amount of energy and happens frequently. Second, Xe is heavy, which is beneficial for thrust generation. Third, Xe is chemically inert and the interaction with the components of the thruster is minimized extending its lifetime. However, Xe is very scarce in the atmosphere, and the separation from air requires many stages of refrigeration and expansion. A production dedicated to Xe would be very costly and not feasible, and usually, it is obtained as a byproduct of oxygen production for industrial applications. Thus, the production of Xe is driven by the demand for industrial oxygen. EP is growing and substituting the traditional chemical propulsion, and the current offer of Xe will not be able to satisfy the foreseen increasing demand \cite{pari10a,holste15}.

The search for candidates for alternative propellants is a current topic of interest. Traditionally, other noble gases, e.g argon and krypton, have been proposed \cite{gian16,vinci21a}. These are more abundant than Xe in the atmosphere but are still rare gases that are produced in the same way as Xe, therefore driven by the industrial oxygen demand. More accessible non-noble substances have been looked into, and several candidates have been found with potential advantages (iodine, water...) \cite{szab13b,bello22,naka18b,molo19}. Some of these candidates allow in-situ resource utilization, very important for in-space missions, and in particular the air-breathing propulsion concept \cite{seme95,cara07,ande18a,hrub22,roma20}: an intake collects and compresses the ambient gases, and after that, a plasma thruster uses the collected gases as propellant. 
These systems have potential application for drag compensation
in low orbits around Earth (with oxygen and nitrogen as ambient gases) and also on other planets with a dense enough atmosphere such as Mars and Venus (with abundant carbon dioxide). Since no propellant storage is required, this concept would allow long-term missions. 
Given its versatility, the air-breathing concept is being considered for different EP technologies, such as the mature Hall Effect Thruster (HET), and the novel family of electrodeless plasma thrusters (EPTs), which includes the Electron Cyclotron Resonance Thruster and the Helicon Plasma Thruster. For HETs, the concept has been considered since some time ago and several prototypes are being developed \cite{seme95,cara07,ande18a,hrub22}. For EPTs, recently Romano et al started the development of the only prototype for now in the literature \cite{roma20}. Non-noble substances, such as air, are reactive though and limit the lifetime of the components of the thruster, such as metallic electrodes. In this respect, EPTs, lacking  electrodes exposed to the plasma, are more tolerant than HETs or Gridded Ion Thrusters. These prototypes are in the very early stages and limited experimental and theoretical characterization of the plasma discharge and performances is available \cite{ande22b}.

The modeling of plasma discharge with molecular chemistry is necessary for the air-breathing propellants. The chemistry of molecules is much more complex than that of noble gases. Molecules can excite in  rotational or vibrational modes and dissociate into atoms after hitting with the electrons. The interaction of plasma species with the thruster walls is more complex too. The dissociated atoms, after hitting the walls, can be recombined into more stable molecular forms via associative wall recombination. Most of the existing models for the plasma discharge in air-breathing thrusters \cite{garr12,tap21,ferr22,souh23}, which treat properly the plasma chemistry, are 0D global. 
Recent progress is due to Taccogna, Cichocki and Minelli \cite{tacc22}, who 
considered the most relevant  chemistry of the main air components 
and implemented it in a multi-dimensional full particle-in-cell (PIC) code, and 
assessed the air-breathing concept in a low power HET. 
Specifically, the HET operated with a nominal discharge voltage of 200V and a mass flow of 0.1mg/s (yielding a discharge power between 50-100W). 
Simulations were run injecting separately N$_2$, O$_2$, and mixtures of N$_2$/O.
Results show the performances with air-breathing propellants are worse than those with Xe.
On the experimental side, HET5k, a 5kW HET prototype, 
was operated with mixtures of N$_2$/O$_2$ finding that performances improve as the  discharge power increases \cite{ferr22}. Regarding EPTs, currently no multi-dimensional simulations have been run, and the studies for the prototype being developed by Romano et al are focused mainly on the system components design \cite{roma20,roma21a}. Recently a first ignition has been achieved, but no  performance measurements are available \cite{roma21b}.

The present work assesses the air-breathing concept for an EPT.
The same chemical reactions and data on the collisional cross sections than
Ref. \cite{tacc22} are used. 
Instead, simulations are run with the hybrid, 2D axisymmetric code HYPHEN
\cite{zhou19a,pera22b,svil21a,zhou22a}. This applies a PIC formulation for heavy species and a fluid formulation for magnetized electrons. This requires to convert the particle-based data on electron-related collisions into fluid-based data for the electron model.
The hybrid code is a good trade-off between reliability and computational cost, and makes parametric studies much more affordable.
The virtual EPT configuration of Ref. \cite{zhou22a} operating on Xe is taken as reference.
Simulations are run injecting N$_2$ and O, and other compositions of interest, in a
power range between 100 and 3000W looking for the point of maximum thrust efficiency. 

The rest of the work is organized as follows. Section \ref{sec:chemistry} describes the plasma chemistry of air components. Section \ref{sec:model} describes the basics of HYPHEN, the plasma discharge model, and the simulation set-up. Section \ref{sec:perf} discusses the performances of the EPT with air components and xenon. Section \ref{sec:2Dmaps} discusses the 2D plasma response. Section \ref{sec:conc} summarizes the conclusions.

\section{Plasma chemistry for air components}
\label{sec:chemistry}

\begin{table}[H]
\begin{center}\resizebox{15cm}{!}{
\begin{tabular}{l  l  l  l c}
\specialrule{.2em}{.1em}{.1em}
\textbf{Number} & \textbf{Label}  &\textbf{Collision} & \boldmath{$\varepsilon_{th}$ [eV]}  & \textbf{Reference}\\
\specialrule{.2em}{.1em}{.1em}

1 & ion01\_N$_2$ & e+N$_2$($\mathrm{X^1} \mathrm{\Sigma_g^+}$) $\rightarrow$ 2e+N$_2^+$  & 15.60 & \cite{tabata06,tabata12} \\ 
2 & exc1\_N$_2$ & e+N$_2$($\mathrm{X^1} \mathrm{\Sigma_g^+}$) $\rightarrow$ e+N$_2$($\mathrm{A^3} \mathrm{\Sigma_u^+}$)  & 6.17 & \cite{tabata06,tabata12} \\ 

3 & exc2\_N$_2$ & e+N$_2$($\mathrm{X^1} \mathrm{\Sigma_g^+}$) $\rightarrow$ e+N$_2$($\mathrm{B^3} \mathrm{\Pi_g}$)  & 7.35 & \cite{tabata06,tabata12} \\ 

4 & exc3\_N$_2$ & e+N$_2$($\mathrm{X^1} \mathrm{\Sigma_g^+}$) $\rightarrow$ e+N$_2$($\mathrm{W^3} \mathrm{\Delta_u}$)  & 7.36 & \cite{tabata06,tabata12} \\ 

5 & exc4\_N$_2$ & e+N$_2$($\mathrm{X^1} \mathrm{\Sigma_g^+}$) $\rightarrow$ e+N$_2$($\mathrm{B'^3} \mathrm{\Sigma_u^-}$)  & 8.16 & \cite{tabata06,tabata12} \\ 

6 & exc5\_N$_2$ & e+N$_2$($\mathrm{X^1} \mathrm{\Sigma_g^+}$) $\rightarrow$ e+N$_2$($\mathrm{a'^1} \mathrm{\Sigma_u^-}$)  & 8.40 & \cite{tabata06,tabata12} \\ 

7 & exc6\_N$_2$ & e+N$_2$($\mathrm{X^1} \mathrm{\Sigma_g^+}$) $\rightarrow$ e+N$_2$($\mathrm{a^1} \mathrm{\Pi_g}$)  & 8.55 & \cite{tabata06,tabata12} \\ 

8 & exc7\_N$_2$ & e+N$_2$($\mathrm{X^1} \mathrm{\Sigma_g^+}$) $\rightarrow$ e+N$_2$($\mathrm{w^1} \mathrm{\Delta_u}$)  & 8.89 & \cite{tabata06,tabata12} \\ 

9-15 & exc8-14\_N$_2$ & e+N$_2$($\mathrm{X^1} \mathrm{\Sigma_g^+}$) $\rightarrow$ e+N$_2^*$  & 11.00-12.90 & \cite{tabata06,tabata12} \\ 








16 & ela\_N$_2$ & e+N$_2$($\mathrm{X^1} \mathrm{\Sigma_g^+}$) $\rightarrow$ e+N$_2$($\mathrm{X^1} \mathrm{\Sigma_g^+}$)  & 0 & \cite{tabata06,tabata12} \\

17 & diss\_N$_2$ & e+N$_2$($\mathrm{X^1} \mathrm{\Sigma_g^+}$) $\rightarrow$ e+N+N  & 9.76 & \cite{tabata06,tabata12} \\

18 & dion\_N$_2$ & e+N$_2$($\mathrm{X^1} \mathrm{\Sigma_g^+}$) $\rightarrow$ 2e+N+N$^+$  & 18.00 & \cite{tabata06,tabata12} \\

19 & rot\_N$_2$ & e+N$_2$($\mathrm{X^1} \mathrm{\Sigma_g^+}$, $J=$ 0) $\rightarrow$ e+N$_2$($\mathrm{X^1} \mathrm{\Sigma_g^+}$, $J=$ All)  & 0 & \cite{trinitiLXCAT} \\ 

20 & vib\_N$_2$ & e+N$_2$($\mathrm{X^1} \mathrm{\Sigma_g^+}$, $\nu=$ 0) $\rightarrow$ e+N$_2$($\mathrm{X^1} \mathrm{\Sigma_g^+}$, $\nu=$ All)  & 0.29 & \cite{tabata06,tabata12} \\

\hline 
 
21&ion01\_N & e+N($\mathrm{^4S}$) $\rightarrow$ 2e+N$^+$  & 14.80 & \cite{bsrLXCAT} \\

22&exc1\_N & e+N($\mathrm{^4S}$) $\rightarrow$ e+N($\mathrm{^2D}$)  & 3.20 & \cite{bsrLXCAT} \\ 

23&exc2\_N & e+N($\mathrm{^4S}$) $\rightarrow$ e+N($\mathrm{^2P}$)  & 4.00 & \cite{bsrLXCAT} \\ 

24-47&exc3-26\_N & e+N($\mathrm{^4S}$) $\rightarrow$ e+N$^*$  & $>$10 & \cite{bsrLXCAT} \\ 

48&ela\_N & e+N($\mathrm{^4S}$) $\rightarrow$ e+N($\mathrm{^4S}$)  & 0 & \cite{bsrLXCAT} \\

\specialrule{.2em}{.1em}{.1em}
\end{tabular}}
\caption{Collisions between electrons and nitrogen considered in the simulations.}
\label{tab:colldata_N2}
\end{center}
\end{table}

\begin{table}[H]
\begin{center}\resizebox{15cm}{!}{
\begin{tabular}{l l  l  l c}
\specialrule{.2em}{.1em}{.1em}
\textbf{Number}  & \textbf{Label}  &\textbf{Collision} & \boldmath{$\varepsilon_{th}$ [eV]}  & \textbf{Reference}\\
\specialrule{.2em}{.1em}{.1em}

1&ion01\_O$_2$ & e+O$_2$($\mathrm{X^3} \mathrm{\Sigma_g^-}$) $\rightarrow$ 2e+O$_2^+$  & 12.10 & \cite{istLXCAT} \\

2&exc1\_O$_2$ & e+O$_2$($\mathrm{X^3} \mathrm{\Sigma_g^-}$) $\rightarrow$ e+O$_2$($\mathrm{a^1} \mathrm{\Delta_g}$)  & 1.00 & \cite{istLXCAT} \\ 

3&exc2\_O$_2$ & e+O$_2$($\mathrm{X^3} \mathrm{\Sigma_g^-}$) $\rightarrow$ e+O$_2$($\mathrm{b^1} \mathrm{\Sigma_g^+}$)  & 1.50 & \cite{istLXCAT} \\ 

4&exc3\_O$_2$ & e+O$_2$($\mathrm{X^3} \mathrm{\Sigma_g^-}$) $\rightarrow$ e+O$_2$($\mathrm{A^3} \mathrm{\Sigma_u^+}$)  & 4.50 & \cite{itik09} \\

5&exc4\_O$_2$ & e+O$_2$($\mathrm{X^3} \mathrm{\Sigma_g^-}$) $\rightarrow$ e+O$_2$($\mathrm{B^3} \mathrm{\Sigma_u^-}$)  & 7.10 & \cite{itik09} \\

6&ela\_O$_2$ & e+O$_2$($\mathrm{X^3} \mathrm{\Sigma_g^-}$) $\rightarrow$ e+O$_2$($\mathrm{X^3} \mathrm{\Sigma_g^-}$)  & 0 & \cite{biagiLXCAT} \\

7&diss1\_O$_2$ & e+O$_2$($\mathrm{X^3} \mathrm{\Sigma_g^-}$) $\rightarrow$ e+O($\mathrm{^3P}$)+O($\mathrm{^3P}$)  & 6.12 & \cite{lawton78} \\

8&diss2\_O$_2$ & e+O$_2$($\mathrm{X^3} \mathrm{\Sigma_g^-}$) $\rightarrow$ e+O($\mathrm{^3P}$)+O($\mathrm{^1D}$)  & 8.40 & \cite{lawton78} \\

9&diss3\_O$_2$ & e+O$_2$($\mathrm{X^3} \mathrm{\Sigma_g^-}$) $\rightarrow$ e+O($\mathrm{^1D}$)+O($\mathrm{^1D}$)  & 9.97 & \cite{lawton78} \\

10&dion\_O$_2$ & e+O$_2$($\mathrm{X^3} \mathrm{\Sigma_g^-}$) $\rightarrow$ 2e+O+O$^+$  & 23.00 & \cite{itik09} \\

11&rot\_O$_2$ & e+O$_2$($\mathrm{X^3} \mathrm{\Sigma_g^-}$, $J=$ 0) $\rightarrow$ e+O$_2$($\mathrm{X^3} \mathrm{\Sigma_g^-}$, $J=$ All)  & 0 & \cite{trinitiLXCAT} \\

12&vib\_O$_2$ & e+O$_2$($\mathrm{X^3} \mathrm{\Sigma_g^-}$, $\nu=$ 0) $\rightarrow$ e+O$_2$($\mathrm{X^3} \mathrm{\Sigma_g^-}$, $\nu=$ All)  & 4.00 & \cite{istLXCAT} \\

\hline

13&ion01\_O & e+O($\mathrm{^3P}$) $\rightarrow$ 2e+O$^+$  & 13.60 & \cite{istLXCAT} \\ 

14&exc1\_O & e+O($\mathrm{^3P}$) $\rightarrow$ e+O($\mathrm{^1D}$)  & 1.97 & \cite{istLXCAT} \\

15&exc2\_O & e+O($\mathrm{^3P}$) $\rightarrow$ e+O($\mathrm{^1S}$)  & 4.19 & \cite{istLXCAT} \\ 

16&exc3\_O & e+O($\mathrm{^3P}$) $\rightarrow$ e+O($\mathrm{^3S^0}$)  & 9.52 & \cite{laher90} \\ 

17-22&exc4-9\_O
& e+O($\mathrm{^3P}$) $\rightarrow$ e+O$^*$  & $>$12 & \cite{laher90} \\ 
23&ela\_O & e+O($\mathrm{^3P}$) $\rightarrow$ e+O($\mathrm{^3P}$)  & 0 & \cite{istLXCAT} \\

\specialrule{.2em}{.1em}{.1em}
\end{tabular}}
\caption{Collisions between electrons and oxygen considered in the simulations.}
\label{tab:colldata_O2}
\end{center}
\end{table}

\begin{table}[H]
\begin{center}\resizebox{15cm}{!}{
\begin{tabular}{l l  l  l c}
\specialrule{.2em}{.1em}{.1em}
\textbf{Number} & \textbf{Label}  &\textbf{Collision} & \boldmath{$\varepsilon_{th}$ [eV]}  & \textbf{Reference}\\
\specialrule{.2em}{.1em}{.1em}

1&ion01\_Xe & e+Xe($\mathrm{5p^6}$ $\mathrm{^1S_0}$) $\rightarrow$ 2e+Xe$^+$  & 12.13 & \cite{biagiLXCAT} \\ 


2-15&exc1-14\_Xe & e+Xe($\mathrm{5p^6}$ $\mathrm{^1S_0}$) $\rightarrow$ e+Xe$^*$  & 8.32-11.58 & \cite{haya03} \\
16&ela\_Xe & e+Xe($\mathrm{5p^6}$ $\mathrm{^1S_0}$) $\rightarrow$ e+Xe($\mathrm{5p^6}$ $\mathrm{^1S_0}$)  & 0 & \cite{hayashiLXCAT} \\

\specialrule{.2em}{.1em}{.1em}
\end{tabular}}
\caption{Collisions between electrons and xenon considered in the simulations.}
\label{tab:colldata_Xe}
\end{center}
\end{table}

For the present work, as a first approach, only collisions between electrons and heavy species are going to be simulated. Collisions between heavy species, such as charge exchange ones, should be important in the plasma plume but are neglected for now. For atomic species, ionization, electronic excitation, elastic and Coulomb collisions are considered. 
For the diatomic species, in addition, dissociation and dissociative ionization, rotational and vibrational excitation need to be considered. 
All the electron-induced collisions are not assumed state-selective, i.e. all species are in the ground state and even after suffering any excitation, the decay to the ground state is instantaneous. Such assumption could not be true for metastable electronic excited states and vibrational excited states, which have longer lifetime and could lead to stepwise ionization processes (i.e. ionization of these excited states instead of direct ionization from ground state). In typical EPT plasmas, electron temperatures are in the range 5-20eV \cite{taka19} and this implies that double ionization events
for air components, with threshold energies of 30-40eV, can be neglected.

Tables \ref{tab:colldata_N2}, \ref{tab:colldata_O2} and \ref{tab:colldata_Xe} list, respectively, the electron-induced collisions for the main components of air, 
nitrogen and oxygen, and xenon to be implemented in the simulations;
and the references where experimental and theoretical data on collisional cross sections versus impact energy have been obtained. 
The labels for the different collisions are: elastic collision (ela), ionization (ion), rotational excitation (rot), vibrational excitation (vib), electronic excitation (exc), dissociation (diss), and dissociative ionization (dion). 

In Table \ref{tab:colldata_N2}, the data of cross sections for N$_2$ is mainly obtained from Refs. \cite{tabata06,tabata12}. Ionization data is only for single ionization, which has a threshold energy of 15.60eV. Electronic excitation data is for collisions from ground state $\mathrm{X^1\Sigma_g^+}$ to 14 excited states, with threshold energies between 6.17 and 12.30eV: $\mathrm{A^3\Sigma_u^+}$, $\mathrm{B^3\Pi_g}$, $\mathrm{W^3\Delta_u}$, $\mathrm{B'^3\Sigma_u^-}$, $\mathrm{a'^1\Sigma_u^-}$, $\mathrm{a^1\Pi_g}$, $\mathrm{w^1\Delta_u}$, $\mathrm{b^1\Pi_u}$, $\mathrm{b'^1\Sigma_u^+}$, $\mathrm{c^1\Pi_u}$, $\mathrm{c_4'^1 \Sigma_u^+}$, $\mathrm{C^3 \Pi_u}$, $\mathrm{E^3 \Sigma_g^+}$ and $\mathrm{a''^1 \Sigma_g^+}$. Elastic collision data is a cross section accounting for the momentum transfer. Dissociation data, with a threshold energy of 9.76eV, accounts for direct dissociation and other additional sources: resonant dissociation, via intermediate unstable negative nitrogen ions; and auto-dissociation, via intermediate unstable nitrogen excited states. Dissociative single ionization, with a threshold energy of 18.00eV, is also considered. Vibrational excitation data is a total cross section from the ground state to all possible final states. Rotational excitation has a negligible threshold energy, so they are treated as elastic collisions for electrons, and a total cross section is used. Data in Refs. \cite{tabata06,tabata12} for rotational excitation is not complete, and instead is taken from Ref. \cite{trinitiLXCAT}. The data for N is obtained from Ref. \cite{bsrLXCAT}. It includes single ionization at 14.54eV; electronic excitation from ground state $\mathrm{^4S}$ to the low energy states $\mathrm{^2D}$ at 2.39eV and $\mathrm{^2P}$ at 3.57eV, and to other 24 states at energies larger than 10eV; and elastic momentum transfer collision. 

In Table \ref{tab:colldata_O2}, the data of cross sections for O$_2$ and O is collected from many sources  \cite{trinitiLXCAT,biagiLXCAT,istLXCAT,itik09,lawton78,laher90} for the same collisions as for nitrogen. For O$_2$, single ionization data, with threshold 12.10eV, is from Ref. \cite{istLXCAT}. Electronic excitation data, from Refs. \cite{istLXCAT,itik09}, is for collisions from ground state $\mathrm{X^3\Sigma_g^-}$ to 4 excited states with thresholds between 1.00 and 7.10eV: $\mathrm{a^1\Delta_g}$, $\mathrm{b^1\Sigma_g^+}$ and $\mathrm{A^3\Sigma_u^+}$ and $\mathrm{B^3\Sigma_u^-}$. Elastic collision data is from Ref. \cite{biagiLXCAT}. The dissociation processes of O$_2$ considered is \cite{lawton78}: 
into oxygen atoms at ground state $\mathrm{^3P}$, at 6.12eV; 
with an atom at ground state $\mathrm{^3P}$ and an atom at excited state $\mathrm{^1D}$, at energy 8.40eV; 
and with both atoms at excited state $\mathrm{^1D}$, at energy 9.97eV. 
Dissociative single ionization data, at energy 23.00eV, is from Ref. \cite{itik09}. Rotational excitation and vibrational excitation use total cross sections from, respectively, Refs. \cite{trinitiLXCAT} and \cite{istLXCAT}. For O, single ionization data is from Ref. \cite{istLXCAT}. Electronic excitation data is from Refs. \cite{istLXCAT,laher90} and considers the collisions from ground state $\mathrm{^3P}$ to 3 low energy states: $\mathrm{^1D}$ at 1.97eV,
$\mathrm{^1S}$ at 4.19eV and $\mathrm{^3S^0}$ at 9.52eV; and to 6 states corresponding to $\mathrm{^5S^0}$, $\mathrm{^3P^0}$, $\mathrm{^5P}$, $\mathrm{^3P}$, $\mathrm{^5D^0}$ and $\mathrm{^3D^0}$ with energies larger than 12eV. Elastic collision data is from Ref. \cite{istLXCAT}.

The data for Xe in Table \ref{tab:colldata_Xe} is from Refs. \cite{biagiLXCAT,hayashiLXCAT,haya03}. 
The collisions simulated are the same as for the atomic air species. Single ionization data is from Ref. \cite{biagiLXCAT}. Electronic excitation data considers the collisions to 14 states with thresholds between 8.32 and 11.58eV collected from Ref. \cite{haya03}. Elastic collision data is collected from Ref. \cite{hayashiLXCAT}.

\begin{figure}[H]
\centering

\begin{minipage}[c]{0.32\textwidth}
\includegraphics[width=1\textwidth]{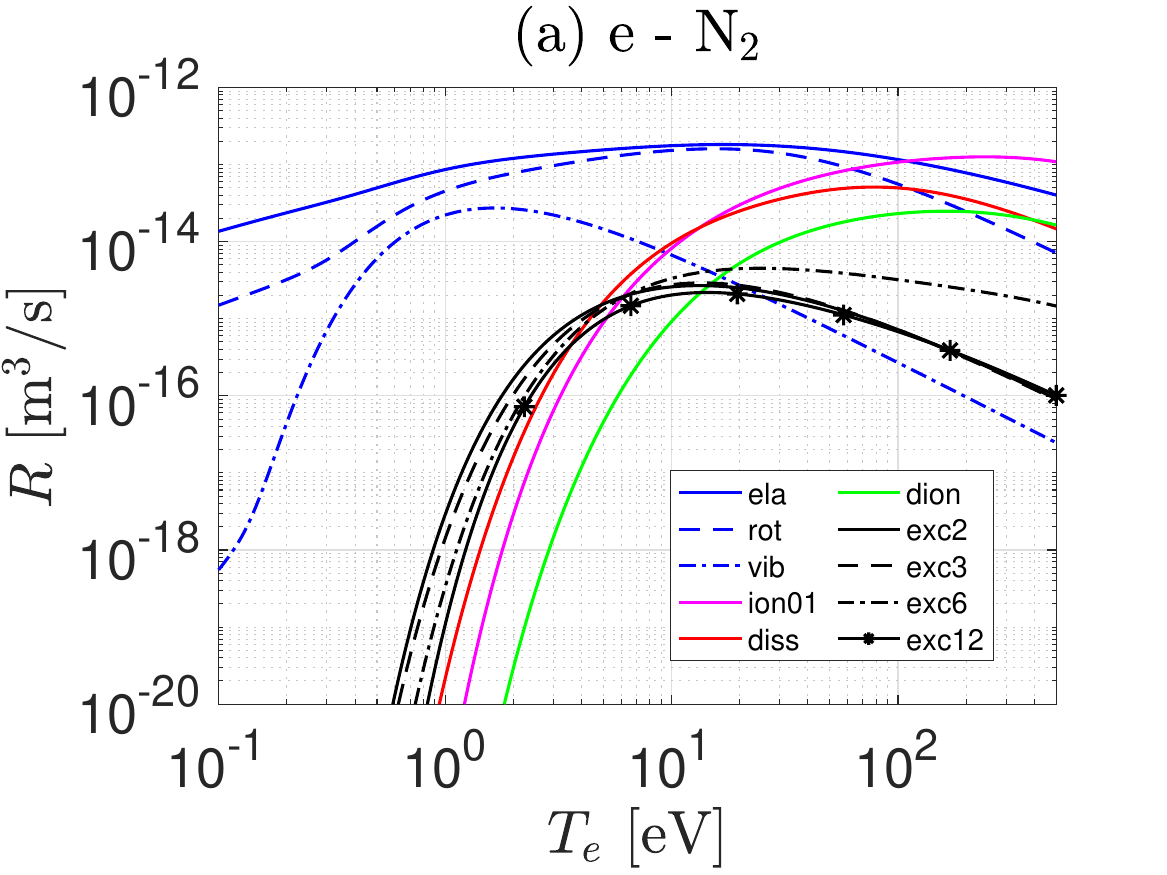}
\end{minipage}
\begin{minipage}[c]{0.32\textwidth}
\includegraphics[width=1\textwidth]{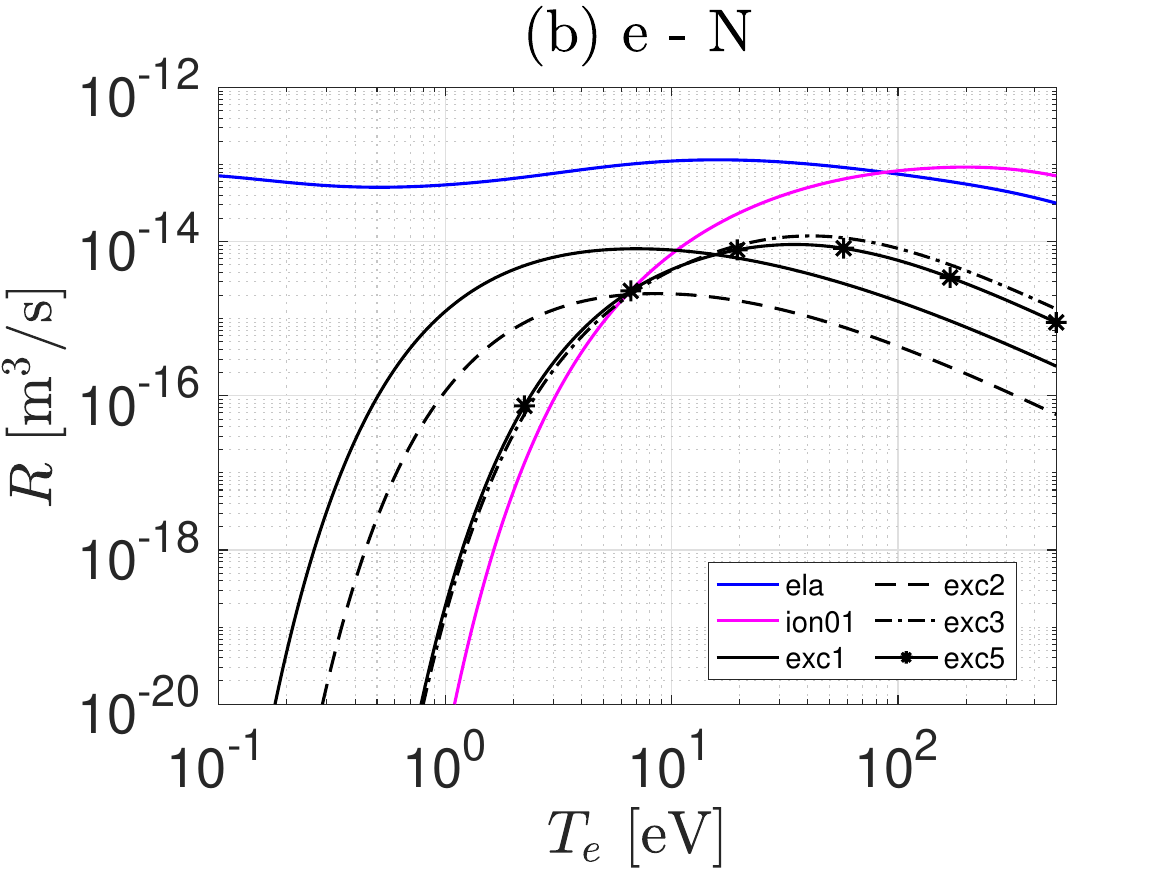}
\end{minipage}

\begin{minipage}[c]{0.32\textwidth}
\includegraphics[width=1\textwidth]{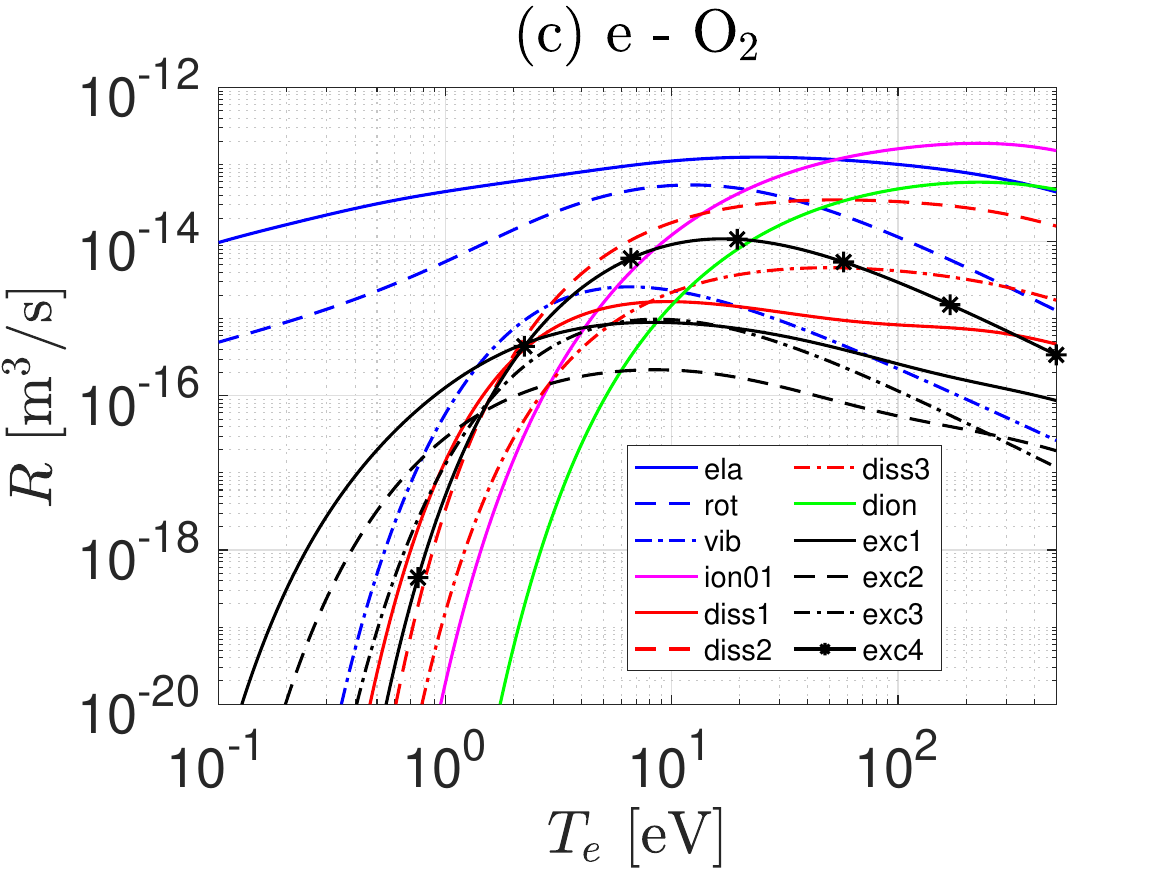}
\end{minipage}
\begin{minipage}[c]{0.32\textwidth}
\includegraphics[width=1\textwidth]{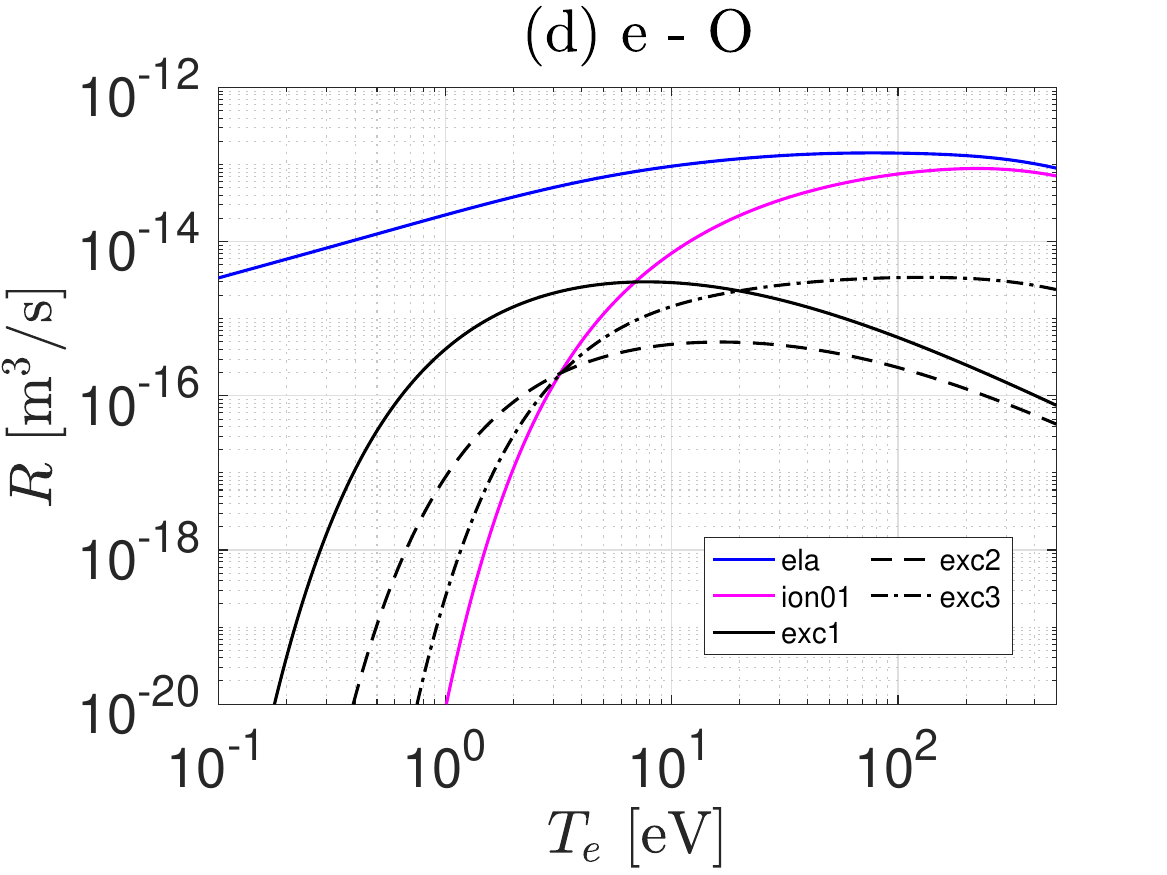}
\end{minipage}

\begin{minipage}[c]{0.32\textwidth}
\includegraphics[width=1\textwidth]{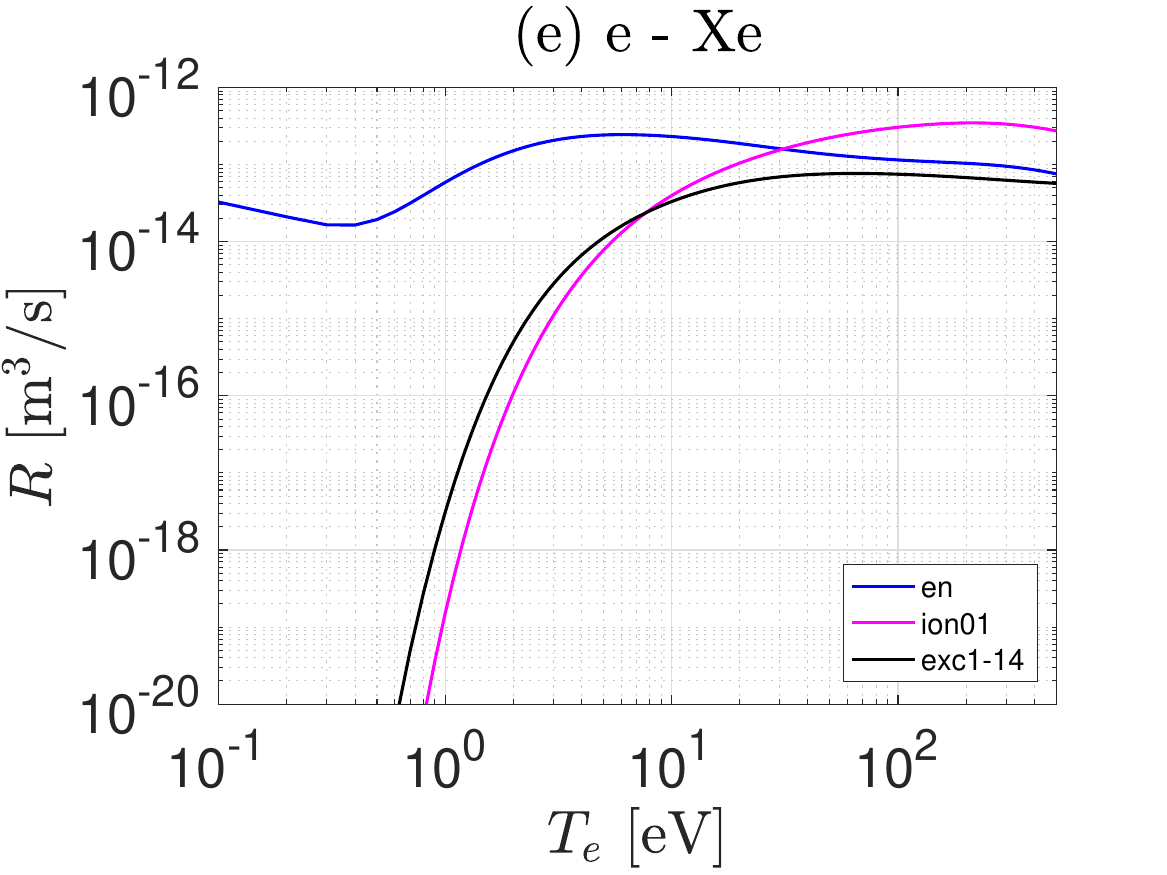}
\end{minipage}
\caption{Rates for collisions of an electron Maxwellian VDF with nitrogen, oxygen, and xenon. (For electronic excitation, rates of some processes are shown, and not all of them.)}
\label{fig:rates}
\end{figure}

In the plasma model to be used next, for collisions between electrons and heavy species, collision rates are needed instead of cross sections. This requires knowing the electron velocity distribution function (VDF).
Assuming a Maxwellian VDF, the collision rate for a collision process $c$ is
\begin{equation}
       R_{c}\left(T_e\right) = \sqrt{\frac{8}{\pi m_eT_e^3}}\int_{\varepsilon_{th,c}}^{\infty}
       \varepsilon\sigma_c(\varepsilon)\exp(-\dfrac{\varepsilon}{T_e})d\varepsilon,
\end{equation}
where $\sigma_c$ and $\varepsilon_{th,c}$ are, respectively, the cross section and threshold energy, $m_e$ is the electron mass and $T_e$ is the electron temperature. Coulomb collisions, which are not shown in the previous tables, 
are also simulated using the well-known analytical formula for its collision rate \cite{BITT04}, 
\begin{equation}
    R_c\left(T_e\right)=k_1T_e^{-3/2}\left(k_2+\frac{1}{2}\ln{\frac{T_e^3}{n_e}}\right);
\end{equation}
where $R_c$ is in m$^3$s$^{-1}$, $T_e$ is in eV, $n_e$ is in m$^{-3}$, and $k_1=2.9\cdot 10^{-12}$ and $k_2=9(1+\ln{10})$. Figure \ref{fig:rates} shows the collision rates for the different collision processes discussed here. The main range of interest here is, say, between 1 and 20eV.

\section{Simulation model}
\label{sec:model}

The simulation model, HYPHEN, has a hybrid formulation. An ion(I)-module applies a PIC approach for the heavy species \cite{domi18c,domi21a}. A dedicated population of (macro)particles is used for each heavy species, and a Cartesian-type mesh, which is defined on the natural cylindrical reference frame \{$\bm{1}_z$, $\bm{1}_r$, $\bm{1}_{\theta}$\}, is used to compute macroscopic quantities of the discrete distribution of macroparticles. An electron(E)-module applies a quasineutral, drift-diffusive (inertialess), magnetized fluid approach for electrons \cite{zhou19a,zhou22a}. To avoid numerical diffusion caused by the large magnetic anisotropy, a 
magnetic field aligned mesh (MFAM) is used, which is defined on the magnetic reference frame \{$\bm{1}_{\parallel}$, $\bm{1}_{\perp}$, $\bm{1}_{\theta}$\}, $\bm{1}_{\parallel}=\bm{B}/B$ and $\bm{1}_{\perp}=-\bm{1}_{\parallel}\times\bm{1}_{\theta}$. An auxiliary sheath model solves the interaction of the thruster surfaces with the electrons at the simulated quasineutral plasma domain. As in Ref. \cite{zhou22a}, the absorption map of electromagnetic power for electrons, from external antennas, is an input for the present simulations. The air-related plasma chemistry defined in the previous section is taken into account in the collisional events and terms of both the PIC and fluid models.

\subsection{PIC model}

The PIC model is based on three algorithms: a particle
mover solves for the trajectories of the macroparticles; a collision operator solves for the collisions of the
macroparticles; and a surface interaction operator solves for the interaction of the macroparticles with
the walls. The plasma chemistry affects the ionization and dissociation processes, and the wall recombination of macroparticles. 
For sake of illustration, let us consider the case of a plasma discharge created from injection of N$_2$.
The collisional processes of Table \ref{tab:colldata_N2} to be considered are (1)-ion01\_N$_2$, (17)-diss\_N$_2$, (18)-dion\_N$_2$ and (21)-ion01\_N.

For the collisions, a Monte Carlo collision method is applied. In a time step $\Delta t$, 
the change of mass of each of the nitrogen species 
in a cell of volume $V_c$, due to ionization and dissociation processes, is the following 
\begin{align}
    &
   \Delta m_{N_2}=-2\left(R_1+R_{17}+R_{18}\right)n_{N_2}n_e  m_N  V_c\Delta t,
   \\
    &
    \Delta m_{N}= \left[\left(2R_{17}+R_{18}\right)n_{N_2}-R_{21} n_N \right] n_e  m_N  V_c\Delta t,
    \\
    &
    \Delta m_{N_2^+}=2 R_1 n_{N_2} n_e  m_N  V_c\Delta t,
    \\
    &
    \Delta m_{N^+}=\left(R_{18} n_{N_2}+R_{21} n_N\right)n_e  m_N  V_c\Delta t,
\end{align}
where $n_s$ is the density of species $s$ (e-electrons, N-atomic nitrogen, N$_2$-molecular nitrogen) and $m_N$ is the mass of atomic nitrogen. If this mass increment is positive, new macroparticles are created with: a position obtained from an uniform sampling inside the cell; a velocity from a Maxwellian VDF sampling given by the colliding species properties; and a certain weight decided by the population control algorithm implemented in the I-module \cite{domi18c}.
If this is negative for a certain species (e.g. $\Delta m_{N_2}$), there is a subtraction proportional to all the macroparticles of that species in the cell.
A similar process is done for O$_2$ and also for Xe (where dissociation processes are not present).

All heavy species impinging a wall are reflected back
with: an accommodated energy $E = (1-\alpha_W)E_{imp}+\alpha_W2T_W$, where $E_{imp}$ is the impact energy of the macroparticle, 
$T_W$
is the temperature of the wall, 
and $\alpha_W$ is the accommodation coefficient; a certain angle defined by the Schamberg model \cite{domi21a}; and 
with the same or different chemical state. Taking again the case of nitrogen for illustration, the following laws are applied:  
N$_2^+$ and N$^+$ macroparticles are first recombined, respectively, into N$_2$ and N; N$_2$ macroparticles are reflected back as N$_2$; a fraction $\gamma$ of N atoms suffers associative wall recombination and are reflected back as N$_2$, and the fraction $(1-\gamma)$ as N. 

\subsection{Electron fluid model}

The electron fluid equations are:
\begin{equation}
    n_e=\sum_{s\ne e}Z_sn_s,
\label{eq:quasi}    
\end{equation}
\begin{equation}
\nabla \cdot \bm{j}_e=-\nabla \cdot \bm{j}_i,
\label{eq:continuity}
\end{equation}
\begin{equation}
\bm 0=-\nabla(n_eT_e)+en_e\nabla \phi+\bm{j}_e\times \bm{B}+\bm{F}_{res}+\bm{F}_{ano},
\label{eq:mom}
\end{equation}
\begin{equation}
    \frac{\partial}{\partial t}\left( \frac{3}{2}n_eT_e\right)+\nabla\cdot \left(\frac{5}{2}T_en_e\bm{u}_e+\bm{q}_e\right)= -\nabla \phi \cdot \bm{j}_e+P'''_a-P'''_{inel},
\label{eq:ene}    
\end{equation}
\begin{equation}
0=-\frac{5n_eT_e}{2e}\nabla T_e-\bm{q}_e\times\bm{B}+\bm{Y}_{res}+\bm{Y}_{ano},
\label{eq:heatflux}
\end{equation}
where $n_e$ is the electron density, $Z_s$ is the charge number; $\bm j_e=- en_e\bm u_e$ is the electron current density and $\bm j_i=\sum_{s\ne e}Z_sen_s\bm u_s$ is the ion current density; $\phi$ is the electric potential; $T_e$ is the electron temperature; and $\bm{q}_e$ is the electron heat flux. The outputs of the equations are $n_e$, $\phi$, $\bm{j}_e$, $T_e$ and $\bm{q}_e$. 

In the current conservation equation \eqref{eq:continuity}, the generation of charge for electrons due to collisions is given by the one for ions according to the quasineutrality and the absence of volumetric sources  \cite{zhou22a}. In the energy conservation equation \eqref{eq:ene}, the power density sink due to collisions is 
\begin{equation}
      P'''_{inel}=n_e\sum_c \nu_{ec}  \varepsilon_{th,c}, 
\end{equation}
where $\nu_{ec}=R_c(T_e)n_{sc}$ is the collision frequency of process $c$ with $n_{s,c}$ the density of the colliding heavy species $s$. Notice that only inelastic collisions (ionization, electronic excitation, vibrational excitation, dissociation and dissociative ionization), with $\varepsilon_{th}>0$, contribute to this sink of power; and for the case of injecting N$_2$, the collisions to be considered are: (1)-ion01\_N$_2$, (2-15)-exc1-14\_N$_2$, (17)-diss\_N$_2$, (18)-dion\_N$_2$, (20)-vib\_N$_2$, (21)-ion01\_N and (22-47)-exc1-26\_N. The term $P'''_a$ is the 
absorbed electromagnetic power density map, which is given as input. If the plasma simulation domain is $\Omega_p$, the absorbed electromagnetic power is $P_a=\int_{\Omega_p} d\Omega P'''_a$ and the inelastic power losses are $P_{inel}=\int_{\Omega_p} d\Omega P'''_{inel}$.

In momentum and heat flux equations \eqref{eq:mom} and \eqref{eq:heatflux}, the collisions introduce resistive terms as 
\begin{equation}
\bm{F}_{res}=m_en_e\sum_c\nu_{ec}\left(\bm{u}_{sc}-\bm{u}_e\right),
    \qquad 
     \bm{Y}_{res}=-\frac{m_e\nu_e}{e}\bm{q}_e,
\end{equation}
where $\bm{u}_{sc}$ is the velocity of the colliding species s, and $\nu_e=\sum_c\nu_{ec}$ is the total collision frequency. The terms $\bm F_{ano}$ and $\bm Y_{ano}$ account for electron anomalous transport and are model phenomenologically as 
\begin{equation}
    \bm{F}_{ano}=\alpha_{ano}Bj_{\theta e}\bm{1}_{\theta},
    \qquad 
    \bm{Y}_{ano}=-\alpha_{ano}Bq_{\theta e}\bm{1}_{\theta}-(m_e\nu_q/e) q_{\parallel e}\bm{1}_{\parallel},
\end{equation}
 with $\alpha_{ano}$ and $\nu_q$ fitting parameters  for cross-field (turbulent) transport and for parallel-field (collisionless) cooling \cite{zhou22a}.

Boundary conditions are imposed locally at the quasineutral edge 
of the Debye sheath next to a wall on
$j_n=\bm{j}\cdot\bm{n}$ and $q_{ne}=\bm{q}_e\cdot\bm{n}$, with $\bm{n}$ the outward unit normal. On the thruster axis, the fluxes are null. 
At the boundary of the finite plume a matching layer with infinity is assumed with $j_n=0$ locally and $q_{ne}$ proportional to the convective energy flux. The sheath model is collisionless and unmagnetized, and accounts for the primary electrons from the plasma bulk and the secondary electron emission (SEE) from the walls (especially significant for dielectric walls). More details on the sheath model and these conditions are given in Ref. \cite{zhou22a}. 

Finite volume methods are applied to the current and energy conservation equations and finite difference methods for the momentum and heat flux equations \cite{zhou19a}. A semi-implicit time scheme is used for temporal derivatives and a number of subiterations is usually applied within a PIC time step \cite{zhou22a}.

\subsection{Simulation set-up}

\begin{figure}[H]
\centering

\begin{minipage}[c]{0.32\textwidth}
\includegraphics[trim = 80mm 40mm 60mm 40mm, clip,width=1\textwidth]{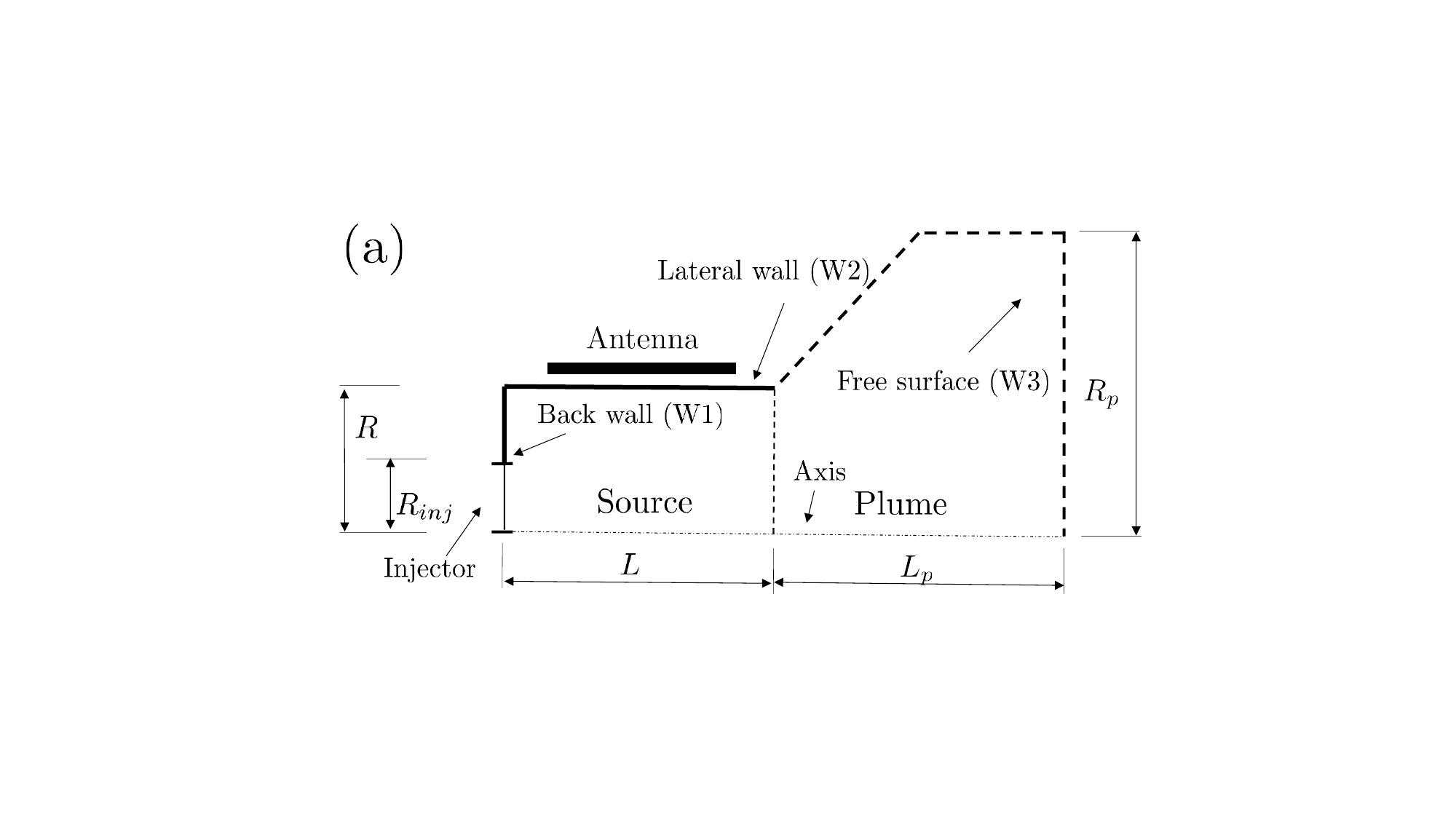}
\end{minipage}
\begin{minipage}[c]{0.32\textwidth}
\includegraphics[trim = 70mm 20mm 70mm 20mm, clip,width=1\textwidth]{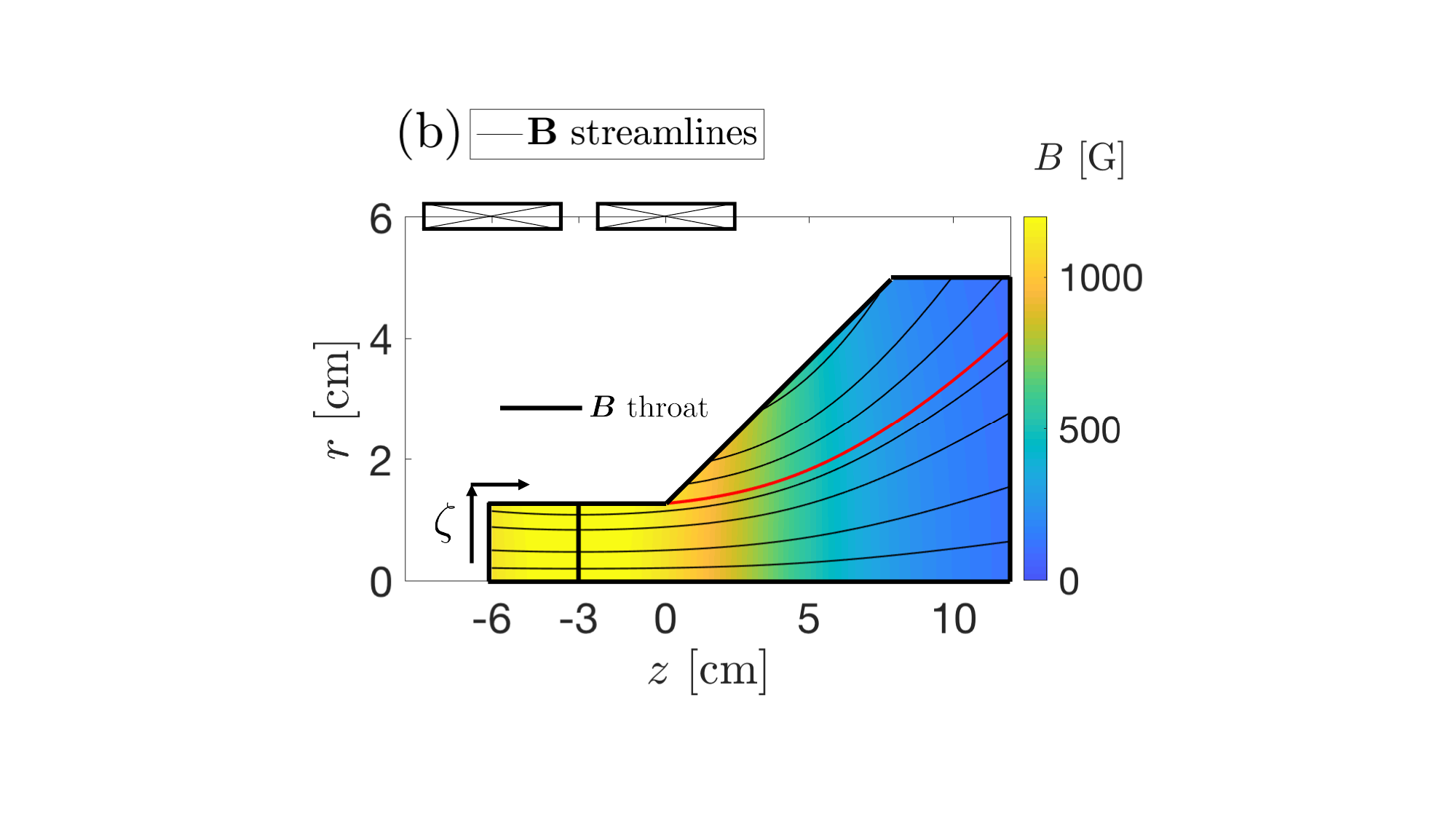} \end{minipage}
\begin{minipage}[c]{0.32\textwidth} 
\includegraphics[width=1\textwidth]{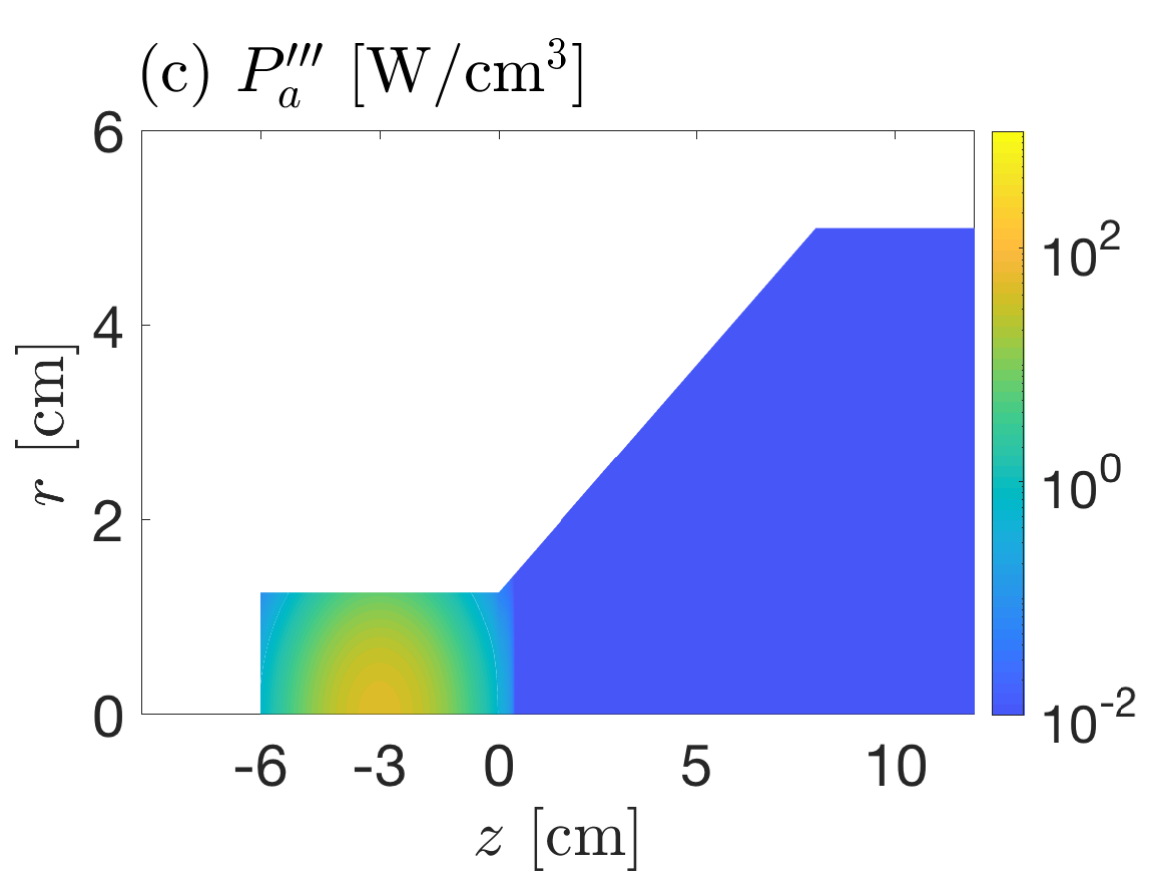}
\end{minipage}

\caption{(a) Sketch of the virtual EPT thruster. (b) Applied magnetic field. (c) Power deposition map.}
\label{fig:thruster_sketch}
\end{figure}

Figure \ref{fig:thruster_sketch}(a)  shows the sketch of the virtual EPT used for the simulations, which is inherited from Ref. \cite{zhou22a} and similar to a prototype under research and designed originally for Xe \cite{nava18a}. The cylindrical channel has a radius $R=1.25$ cm and a length $L=3$ cm; $z=0$ is placed at the channel exit; walls are made of boron nitride.
The topology of the applied magnetic field, Fig. \ref{fig:thruster_sketch}(b), is quasi-axial inside the source and divergent outside in the plume; the maximum strength is about 1200G at the central section of the channel $z=-3$ cm.
In all simulations, the power deposition map $P'''_a$, Fig. \ref{fig:thruster_sketch}(c), is assumed Gaussian and centered at 
$z=-3$ cm too.
The injector, centered at the channel back surface
has radius $R_{inj}=0.625$ cm.
The main composition of the air in the ionosphere, where the air-breathing systems usually operate in space missions, are N$_2$ and O. Simulations are run first independently for N$_2$, O and Xe for various powers (between 10 and 3000W); and a nominal propellant mass flow of $\dot{m}=1$ mg/s is injected for all cases.

The simulation domain is limited by the back wall W1, the lateral wall W2 and the plume free loss surface W3. The simulated plume has a conical shape, with a length $L_p=12$ cm and a maximum radius $R_p=5$ cm. The same meshes as in Ref. \cite{zhou22a} are used, with 1800 cells for the Cartesian mesh of the I-module, and 1961 cells for the MFAM mesh of the E-module. The time step used for ions is $\Delta t=10^{-8}$ s and the number of subiterations within the E-module is $N_e=5$. The total simulated time is 1.5ms, large enough to assure a stationary response.

Table \ref{tab:sim_par} shows the main simulation parameters. 
The anomalous transport parameters, fitted for Xe in Ref. \cite{zhou22a},  are $\alpha_{ano}=0.02$ and $\nu_q=5\cdot10^8$ s$^{-1}$. Due to the lack of experimental data, these values are also used for the air species.  The wall accommodation is assumed total with $\alpha_W=1$, and the associative wall recombination coefficients for air species are selected as $\gamma_{N}=0.07$  and $\gamma_{O}=0.17$ for, respectively, nitrogen and oxygen based on Refs. \cite{thor09,gud13}.

\begin{table}[H]
\begin{center}
\begin{tabular}{ c  c  c c}
\specialrule{.2em}{.1em}{.1em}
\textbf{Simulation parameter}  &\textbf{Symbol} &\textbf{Units}  & \textbf{Value}\\
\specialrule{.2em}{.1em}{.1em}
Thruster length & $L$ & cm & 6 \\ 
Thruster radius & $R$ & cm & 1.25 \\ 
Maximum magnetic strength & - & G & 1200 \\

Power deposition profile & - & - & Gaussian \\

Injector radius & $R_{inj}$& cm & 0.625\\
Total mass flow & $\dot{m}$ & mg/s & 1\\

\hline

Plume length & $L_p$ & cm & 12 \\ 
Maximum plume radius & $R_p$ & cm & 5 \\ 

I-mesh size& - & cells & 1800\\ 
E-mesh size& - & cells & 1961\\ 

I-module time step & $\Delta t$ & s & 10$^{-8}$\\ 
E-module time subiterations & $N_e$ & - & 10\\ 
Total simulation time & - & ms & 1.5 \\

Anomalous transport coefficient & $\alpha_{ano}$ & - & 0.02\\ 
Anomalous cooling collisionality & $\nu_q$ & s$^{-1}$ & 5 $\cdot$ 10$^8$\\ 

Wall accommodation coefficient & $\alpha_W$ & - & 1.00\\ 

Associative wall recombination coefficient (N) & $\gamma_N$ & - & 0.07\\ 

Associative wall recombination coefficient (O) & $\gamma_O$ & - & 0.17\\ 

\specialrule{.2em}{.1em}{.1em}
\end{tabular}
\caption{Simulation parameters common for all the cases.}
\label{tab:sim_par}
\end{center}
\end{table}

\section{Analysis of performances}
\label{sec:perf}

First, several efficiencies helpful for the discussion of the performances of the different propellants and powers are defined.
Following Ref. \cite{zhou22a}, the ion production balance in the steady state is
\begin{equation}
    \dot{m}_{i,total} = \dot{m}_{i,wall}+\dot{m}_{i,beam},
\end{equation}
where $\dot{m}_{i,total}$ is the total ion mass production, $\dot{m}_{i,wall}$ is the ion mass flow to the thruster walls (W1+W2), and $\dot{m}_{i,beam}$ is ion mass flow downstream (across the plume free loss surface W3). The three terms include both molecular and atomic ions. The propellant mass utilization and the ion production efficiency are defined, respectively, as
\begin{equation}
    \eta_u=\dot{m}_{i,beam}/\dot{m},
    \qquad
    \eta_{prod}=\dot{m}_{i,beam}/\dot{m}_{i,total}.
\end{equation}
Summing for all species, the plasma power balance in the steady state is 
\begin{equation}
    P_a =P_{inel}+ P_{wall}+P_{beam},
\end{equation}
where $P_{wall}$ and 
$P_{beam}$ are, respectively, the plasma energy flows to the thruster walls (W1+W2) and downstream (W3). The thrust efficiency is defined and factorized as 
    \begin{equation}
   \eta_F=\frac{F^2}{2\dot{m}P_a}\equiv \eta_{ene}\eta_{div}\eta_{disp},
\end{equation}
with
\begin{equation}
   \eta_{ene}=\frac{P_{beam}}{P_a},
   \qquad
   \eta_{disp} = \frac{F^2}{2\dot{m}P_{beam}^{(z)}},
   \qquad
    \eta_{div} = \frac{P_{beam}^{(z)}}{P_{beam}},
\end{equation}  
respectively, the energy, plume-dispersion and plume-divergence efficiencies. Here $F$ and $P^{(z)}_{beam}$ are the thrust and the plasma axial energy flow downstream.

\begin{table}[H]
\begin{center}\resizebox{\textwidth}{!}{
\begin{tabular}{c c c | c  c   c  c  c  c  c  c  c   }
\specialrule{.2em}{.1em}{.1em}
\textbf{\small  Case} & \textbf{\small Prop.} & \boldmath{$P_a$ [W]} &\boldmath{$F$ [mN]} & \boldmath{$\eta_F$} & \boldmath{$\eta_{ene}$} & \boldmath{$\eta_{disp}$} & \boldmath{$\eta_{div}$}  & \boldmath{$\eta_u$} & \boldmath{$\eta_{prod}$}  & \boldmath{$P_{wall}/P_a$}  &  \boldmath{$P_{inel}/P_a$}\\ 

\specialrule{.2em}{.1em}{.1em}
1&\small Xe & 300  &$7.09$ &  $0.084$  & $0.164$  &  $0.66$ & $0.77$  & $0.95$ &$0.14$  & $0.445$ & $0.391$\\ 

 \hline
2&\small O & 300 &$5.09$ &  $0.043$  & $0.162$  &  $0.32$ & $0.82$  & $0.34$ &$0.38$  & $0.212$ & $0.626$\\ 

3&\small O & 1500  &$19.76$ &  $0.132$  & $0.268$  &  $0.58$ & $0.86$  & $0.97$ &$0.28$  & $0.461$ & $0.271$\\ 

 \hline
4&\small N$_2$ &300 &$3.11$ &  $0.016$  & $0.088$  &  $0.22$ & $0.82$  & $0.20$ &$0.37$  & $0.125$ & $0.787$\\ 

5&\small N$_2$ & 1500  &$14.31$ &  $0.067$  & $0.135$  &  $0.60$ & $0.83$ & $0.82$ &$0.32$  & $0.178$ & $0.687$\\ 

6&\small N$_2$ & 2000  &$19.93$ &  $0.099$  & $0.191$  &  $0.62$ & $0.84$ & $0.97$ &$0.29$  & $0.295$ & $0.514$\\

 \hline

7&\small N$_2$-O & 300  &$3.83$ &  $0.024$  & $0.116$  &  $0.25$ & $0.82$ & $0.25$ &$0.37$  & $0.159$ & $0.725$\\ 

8&\small N$_2$-O & 1500  &$18.41$ &  $0.113$  & $0.214$  &  $0.62$ & $0.85$ & $0.95$ &$0.30$  & $0.332$ & $0.454$\\ 
 \hline

9& \small O$_2$ & 300 &$4.60$ &  $0.035$  & $0.143$ & $0.29$ &  $0.82$  & $0.31$ &$0.36$  & $0.184$ & $0.673$\\ 

10&\small O$_2$ & 1500 &$19.60$ &  $0.126$  & $0.268$ & $0.57$ &  $0.85$  & $0.97$ &$0.28$  & $0.445$ & $0.287$\\ 

\specialrule{.2em}{.1em}{.1em}
\end{tabular}}
\caption{Performances for various propellants and powers with $\dot{m}=$ 1 mg/s (mixture N$_2$/O is 50/50 in mass flow). $F$ in mN is also the specific impulse $F/\dot m$ in km/s.}
\label{tab:perf_ind}
\end{center}
\end{table}

Table \ref{tab:perf_ind} shows the thruster performances for various propellants and powers.  The mass flow is 
$\dot{m}=$ 1 mg/s in all cases; hence, $F$ in mN is also the specific impulse $I_{sp}=F/\dot m$ in km/s. The different cases  are discussed next. 

{\it Case 1: Xe and 300W}. 
This is basically the nominal case of Ref. \cite{zhou22a} except for some updates on the collisional database.
The propellant utilization is excellent, 
$\eta_u=95$\%, but the thrust efficiency is rather modest, $\eta_F=8.4$\%. The main reason is that there is 
too much re-ionization: $\eta_{prod}=14$\% means that a neutral is ionizes around 6-7 times.  
This implies the large fraction of inelastic losses ($P_{inel}/P_a=39.1$\%) and wall losses  ($P_{wall}/P_a=44.5\%$), and as a result the energy efficiency is rather low 
($\eta_{ene}=16.4$\%). 
The dispersion efficiency, which measures the lack of monoenergicity of the downstream beam, is reasonable ($\eta_{disp}=66\%$) and the divergence efficiency ($\eta_{div}=77\%$) corresponds to a divergence semiangle of 30 deg.

{\it Cases 2 and 3: O, for 300W and 1500W}. At 300W, the propellant utilization and thrust efficiencies are very modest: $\eta_u=34$\% and $\eta_F=4.3\%$.
This operation point, with a very large inelastic loss fraction ($P_{inel}/P_a=62.6\%$) and a low electron temperature, seems far from the optimum one for atomic oxygen.
At 1500W,  performances have improved a lot and they are even better than those of Xe at 300W (case 1): $\eta_u=97$\%, $\eta_F=13.2\%$, and
$\eta_{ene}=26.8$\%. Inelastic losses decrease to $P_{inel}/P_a=27.1$\%, while wall losses increase to $P_{wall}/P_a=46.1$\%. The production efficiency at 1500W is 28\%, double than for Xe.
Thus, although Xe is easier to ionize, it is also subjected to more re-ionization, which in the end is detrimental. The dispersion efficiency,  $\eta_{disp}=58\%$, is not high, but justified by the presence of both atoms and molecules. The plume divergence semiangle is about 22deg
(corresponding to $\eta_{div}=86\%$), lower than for Xe.

{\it Cases 4 to 6: N$_2$ for 300W, 1500W and 2000W}. The case of 300W is very inefficient as the case of O at 300W (case 2), even worse (with $\eta_u=20\%$ and $\eta_F=1.6\%$).
At 1500W, performances with N$_2$ are comparable to case 1 but still significantly lower than case 3: $\eta_u=82$\%, $\eta_F=6.7\%$ and $\eta_{ene}=13.5$\%. The inelastic losses are still very large ($P_{inel}/P_a = 68.7\%$), which suggests that the electron temperature is still low.
At 2000W, performances become better than case 1 and comparable with case 3: $\eta_u=97$\%, $\eta_F=9.9\%$ and $\eta_{ene}=19.1$\%. The production efficiency, $29\%$, is similar to case 3 and thus better than for case 1.

\begin{figure}[H]
\centering

\begin{minipage}[c]{0.35\textwidth}
\includegraphics[width=1\textwidth]{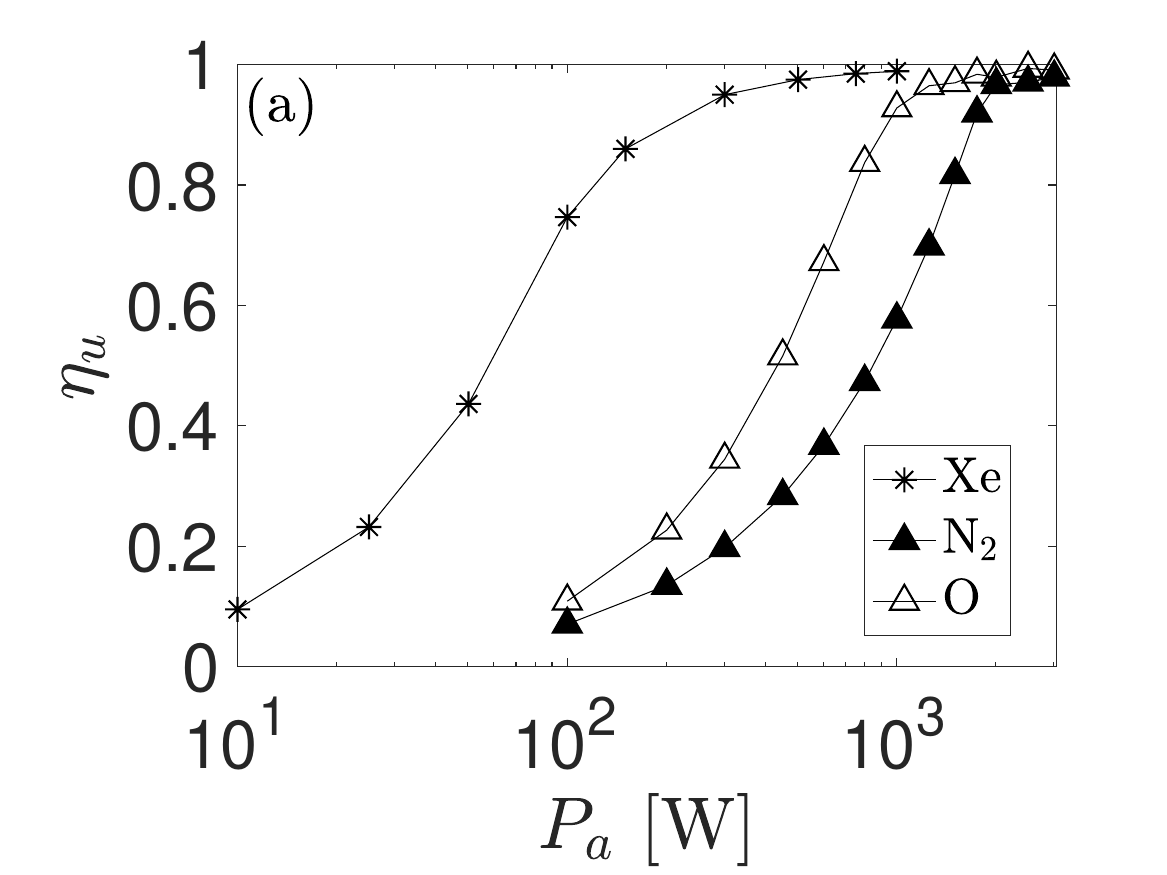}
\end{minipage}
\begin{minipage}[c]{0.35\textwidth}
\includegraphics[width=1\textwidth]{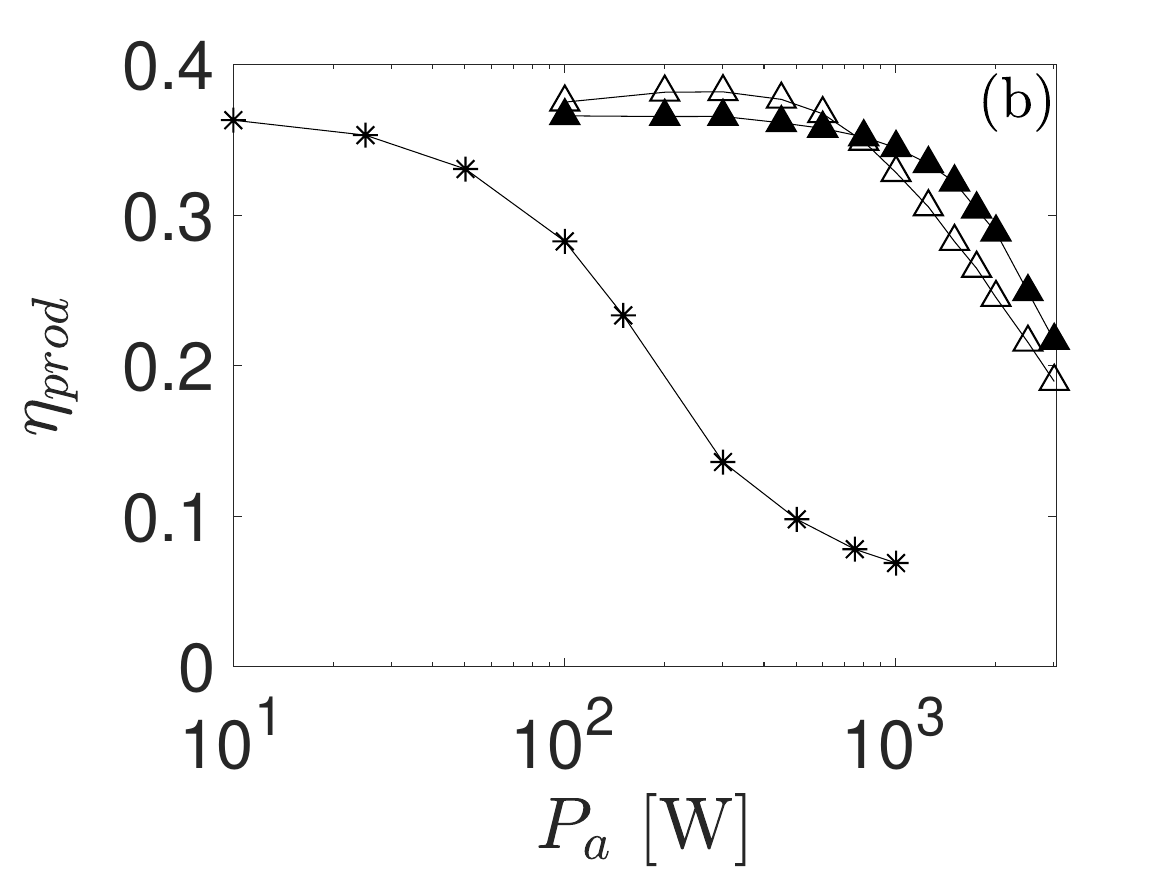}
\end{minipage}

\begin{minipage}[c]{0.35\textwidth}
\includegraphics[width=1\textwidth]{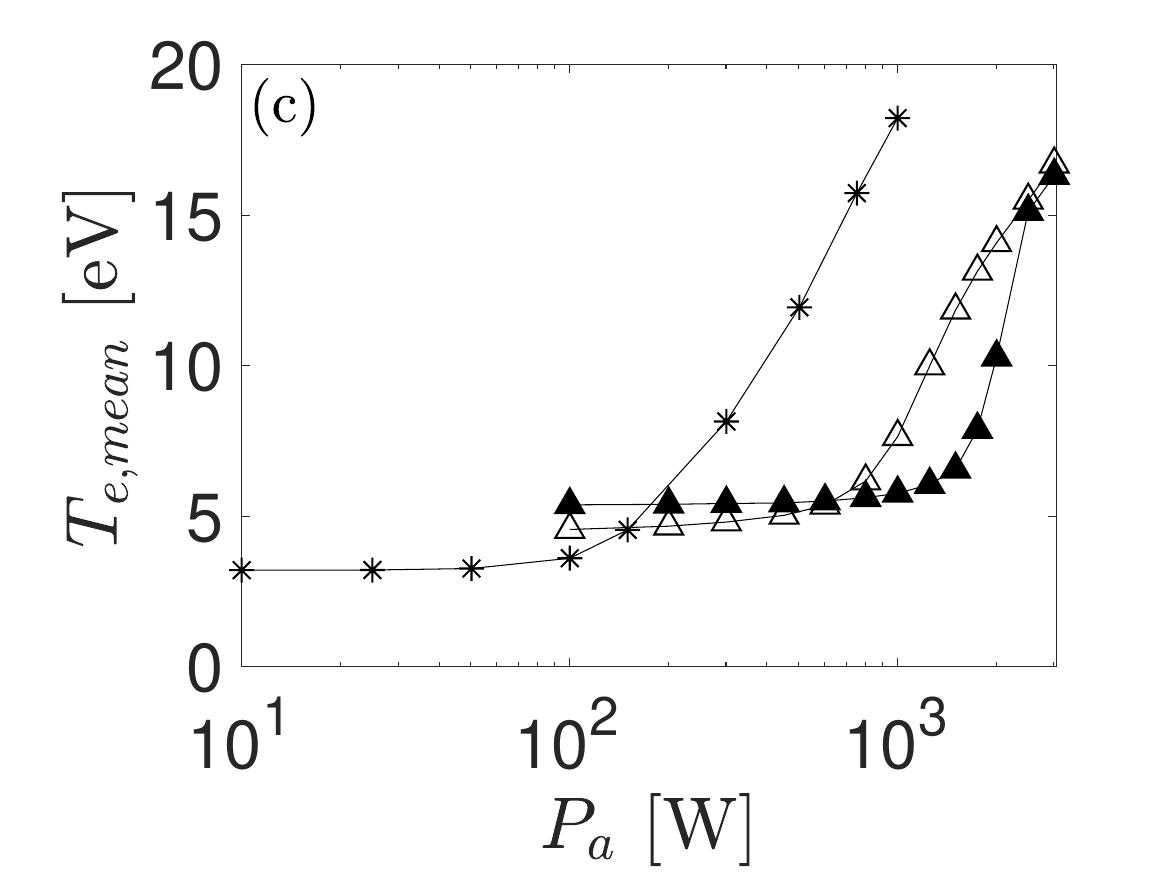}
\end{minipage}
\begin{minipage}[c]{0.35\textwidth}
\includegraphics[width=1\textwidth]{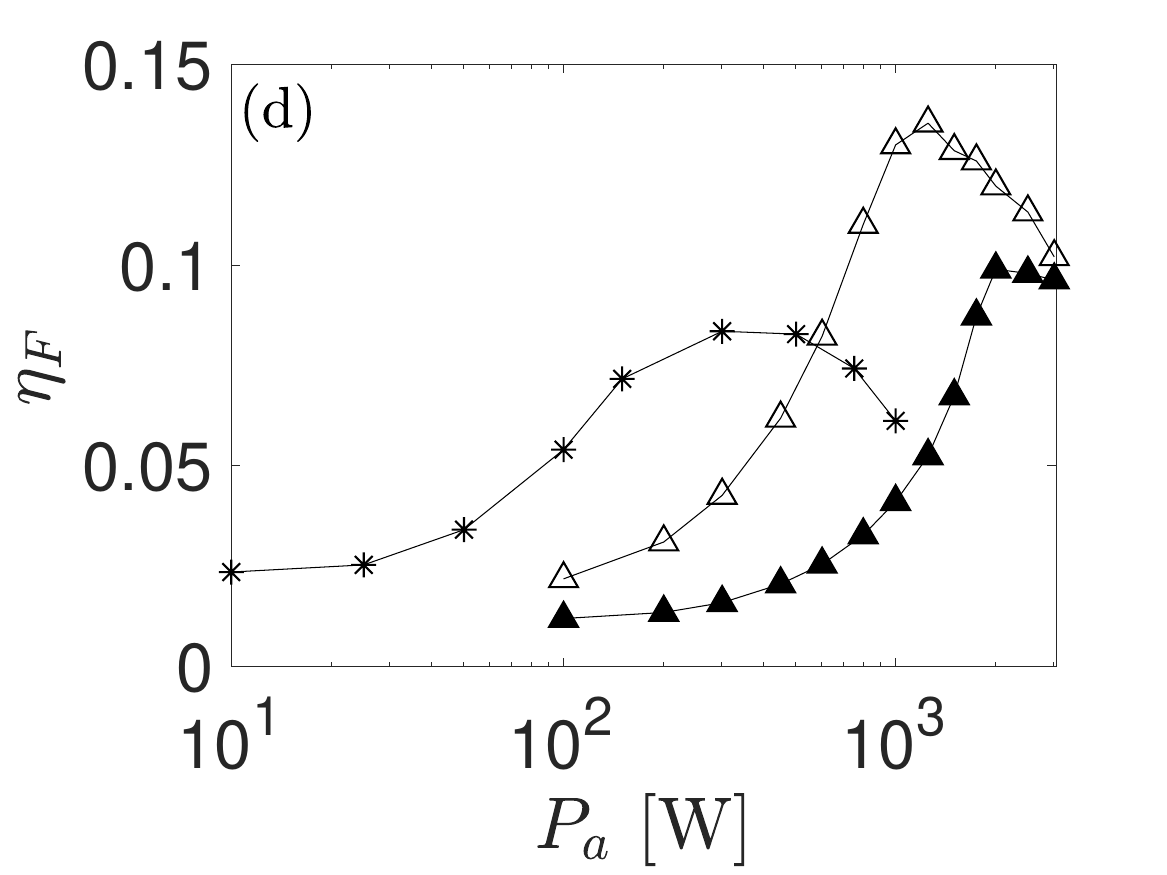}
\end{minipage}

\caption{Thruster performances versus $P_a$ for Xe ($\ast$), O ($\triangle$) and N$_2$ ($\blacktriangle$).
(a) Propellant utilization, (b) production efficiency, (c) mean electron temperature over the plasma source and (d) 
thrust efficiency.
}
\label{fig:vspower}
\end{figure}

Figure \ref{fig:vspower} shows how the performances of the different propellants are related to the power in the range 10-1000W for Xe and 100-3000W for O and N$_2$. Interestingly the trends for Xe, O and N$_2$ are similar, but the transition to full ionization and maximum thrust efficiency happens at different (optimal) powers. For instance, $\eta_u\approx 75\%$ [panel (a)] is reached at about
100W for Xe, 700W for O and 1400W for N$_2$. That value of the propellant utilization corresponds roughly to a power where (i) $\eta_{prod}(P_a)$ [panel (b)] starts to decrease, and (ii) the knees in the curves of the mean source electron temperature $T_{e,mean}(P_a)$ [panel (c)] are located.
The curves $\eta_F(P_a)$ [panel (d)] present a maximum of 8.4\% for Xe, 13.5\% for O and 9.9\% for N$_2$, reached at optimal powers about 300W, 1250W and 2000W respectively. 
The larger maximum thrust efficiency of O and N$_2$ with respect to Xe is remarkable.
The mean temperatures at the maxima of $\eta_F$ are $T_{e,mean}\sim$ 8-10 eV, and at these temperatures, there is nearly full ionization of the propellant and re-ionization becomes a strong source of losses. These common trends for the different propellants indicate a robust plasma response. 
As discussed in Ref. \cite{zhou22a}, improvements in the magnetic confinement of the plasma or reduction of the thruster channel length would increase
$\eta_{prod}$ and therefore $\eta_F$.

Cases 2 to 6, studied so far, inject N$_2$ and O independently. For air-breathing systems in the ionosphere, a mixture of both is taken by the intake. Furthermore, O tends to become O$_2$ at the intake walls via associative recombination, so molecular oxygen is injected into the thruster. Cases 6 to 9 in Table \ref{tab:perf_ind} try to assess the impact of these features.

\begin{figure}[H]
\centering

\begin{minipage}[c]{0.35\textwidth}
\includegraphics[width=1\textwidth]{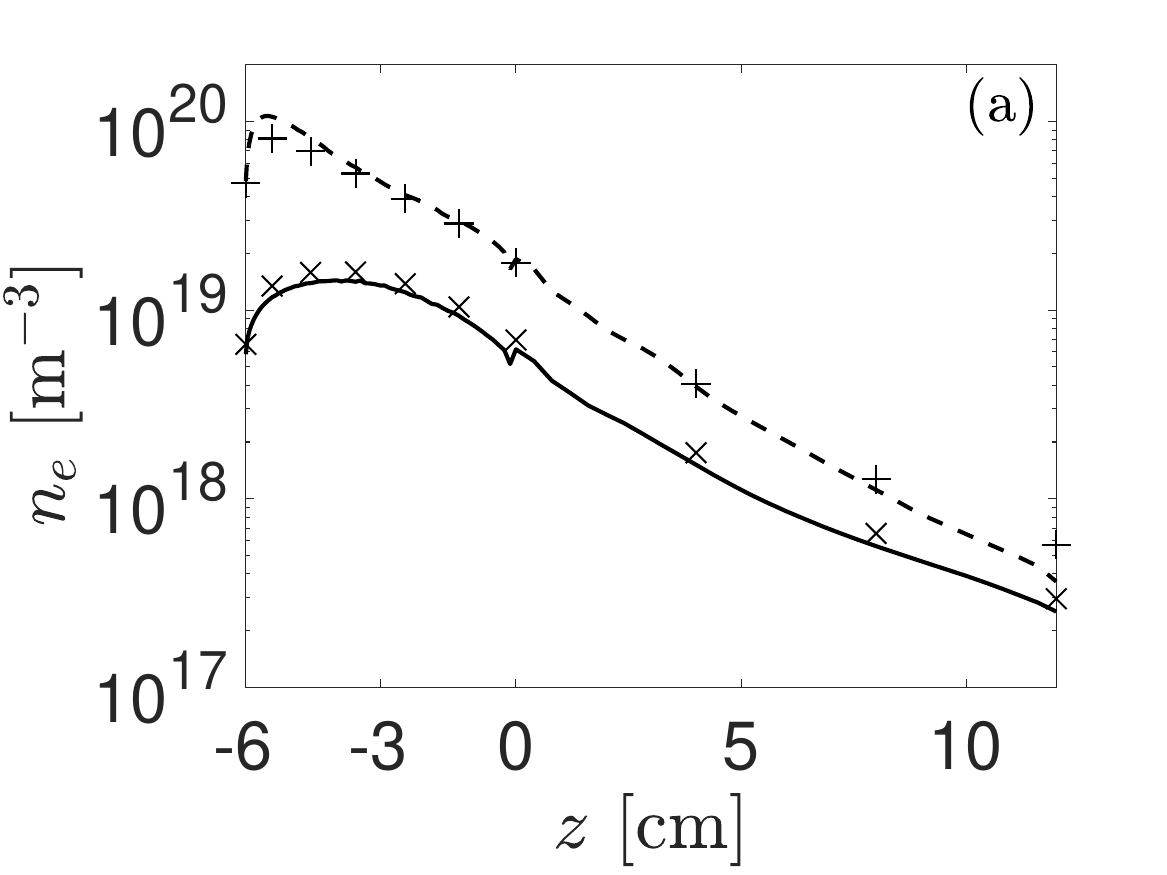}
\end{minipage}
\begin{minipage}[c]{0.35\textwidth}
\includegraphics[width=1\textwidth]{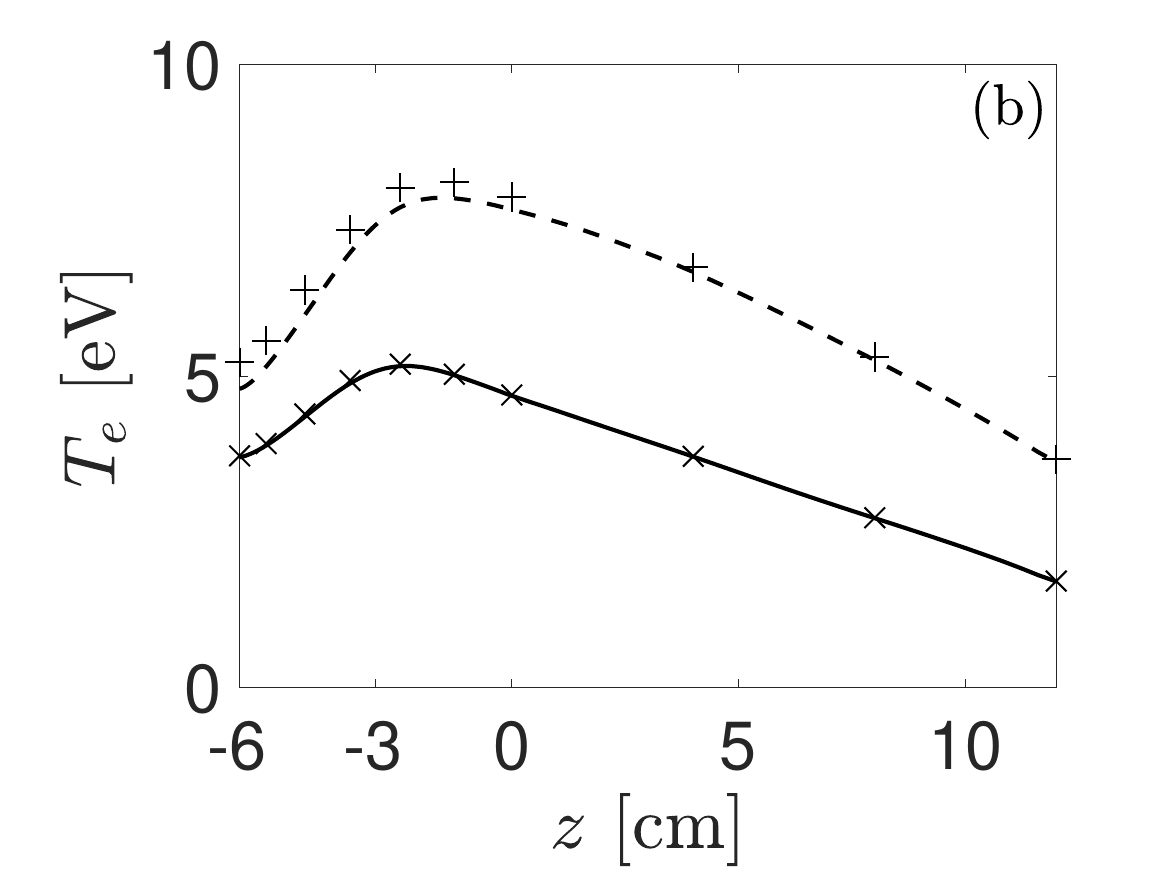}
\end{minipage}

\caption{Injection of a 50/50 mixture of N$_2$/O. Spatial profiles along the axis of (a) electron density and (b) electron temperature at $P_a=$ 300 W (solid lines) and 1500 W (dashed lines). Marks (x) and (+) are fittings points from weighted averages of the independent cases of N$_2$ and O at $P_a=$ 300 W and 1500 W.}
\label{fig:vsprop1}
\end{figure}

{\it Case 7 and 8: N$_2$-O, 300W and 1500W}.
 Simulations are run keeping the total mass flow of $\dot{m}=$ 1mg/s with a mass share of 50\% for N$_2$ and O. 
 Both Table \ref{tab:perf_ind} and Fig. \ref{fig:vsprop1} demonstrate that 
 (as long as collisions between heavy species are negligible) the plasma response operating with the N$_2$/O mixture 
 is well estimated 
 by the  weighted average of the response for the independent propellants. Figure \ref{fig:vsprop1} shows that the differences between the multi-propellant response
 and the fitted one, for electron density $n_e$ and temperature $T_e$, is below 10\%.

\begin{figure}[H]
\centering

\begin{minipage}[c]{0.35\textwidth}
\includegraphics[width=1\textwidth]{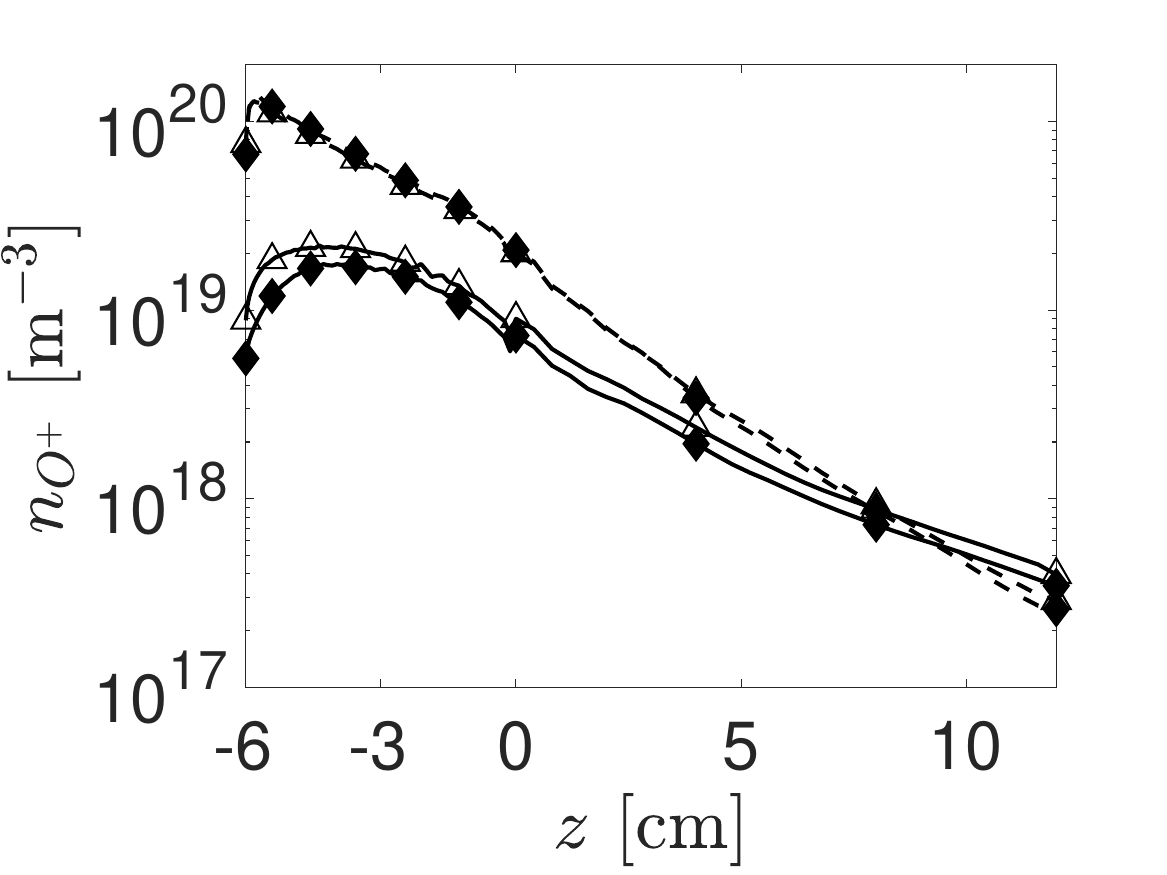}
\end{minipage}

\caption{Injection of O$_2$. Spatial profile along the axis of the atomic ion density injecting O$_2$ ($\blacklozenge$) and O ($\lozenge$) at
$P_a=$ 300 W (solid lines) and 1500 W (dashed lines).}
\label{fig:vsprop2}
\end{figure}

{\it Case 9 and 10: $O_2$, 300W and 1500W}.
Once inside the thruster, O$_2$ dissociation into O is very strong, which explains that 
the plasma response for the cases injecting O and O$_2$ are very similar, mainly at high power.
Figure \ref{fig:vsprop2} shows for $P_a=$ 300 W and 1500 W, the atomic ion density $n_{O^+}$ for injection of O and O$_2$. At 300W, 
the difference is mild, and for the case of interest, 1500W, it is negligible.

The trends found here agree well with the limited numerical studies of Ref. \cite{tacc22} with a low power HET 
and experimental measurements on a 5kW HET of Ref. \cite{ferr22}.
In the first case, for the operation conditions designed for Xe, with a discharge voltage of 200V and a mass flow of 0.1mg/s, performances with air are shown worse, and an incomplete ionization and low electron temperature are found. Thrust and utilization efficiencies for a 50/50 mixture of N$_2$/O are 5\% and 46\%, while for Xe, these values are 10\% and 85\%. In the second case, the 5kW thruster is operated with a mixture of N$_2$/O$_2$, discharge voltages of 225 and 300V, and mass flows of 5, 6 and 7mg/s. Among the few performances studied, the overall efficiency increases with the discharge power reaching a maximum value of 20\%, while  values about 50-60\%  are achieved when using Xe \cite{pera22b}.

\section{2D plasma response}
\label{sec:2Dmaps}

\begin{figure}[H]
\centering

\begin{minipage}[c]{0.36\textwidth}
\includegraphics[width=1\textwidth]{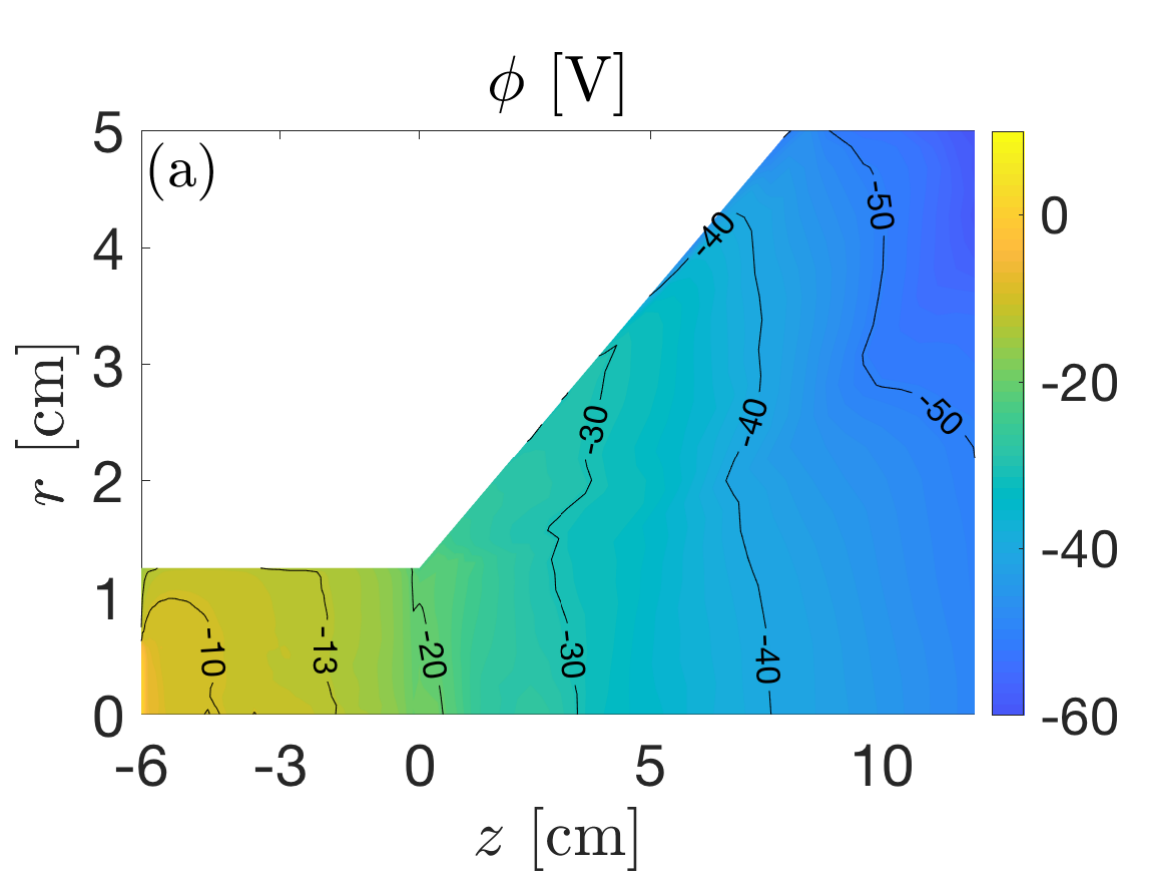}
\end{minipage}
\begin{minipage}[c]{0.36\textwidth}
\includegraphics[width=1\textwidth]{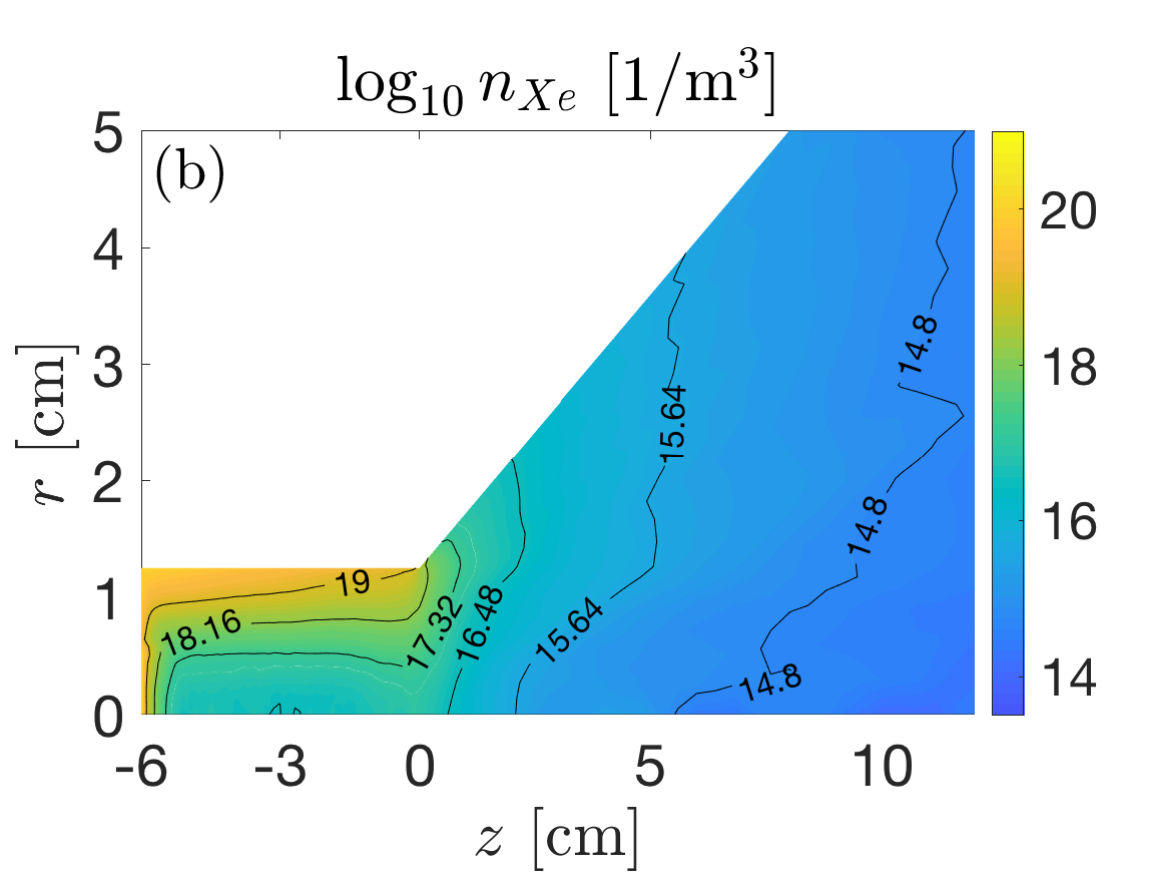}
\end{minipage}

\begin{minipage}[c]{0.36\textwidth}
\includegraphics[width=1\textwidth]{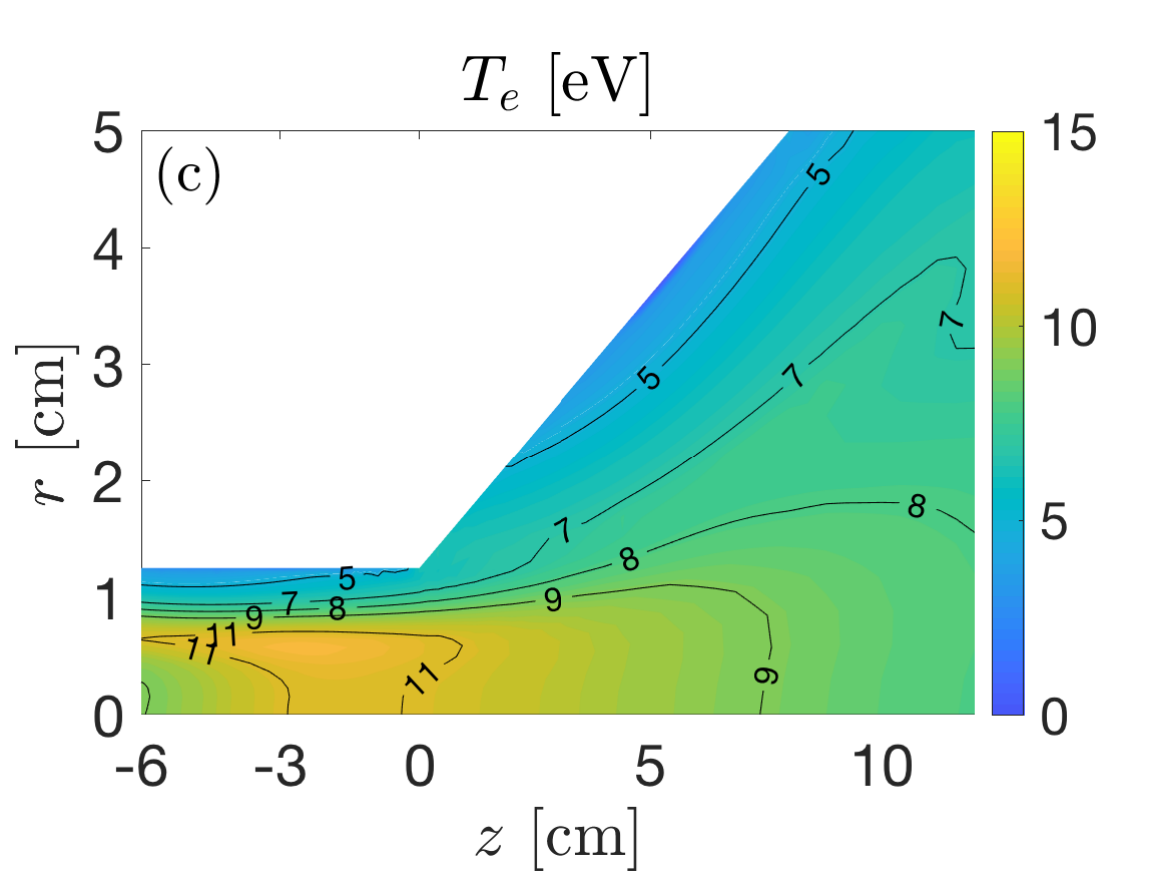}
\end{minipage}
\begin{minipage}[c]{0.36\textwidth}
\includegraphics[width=1\textwidth]{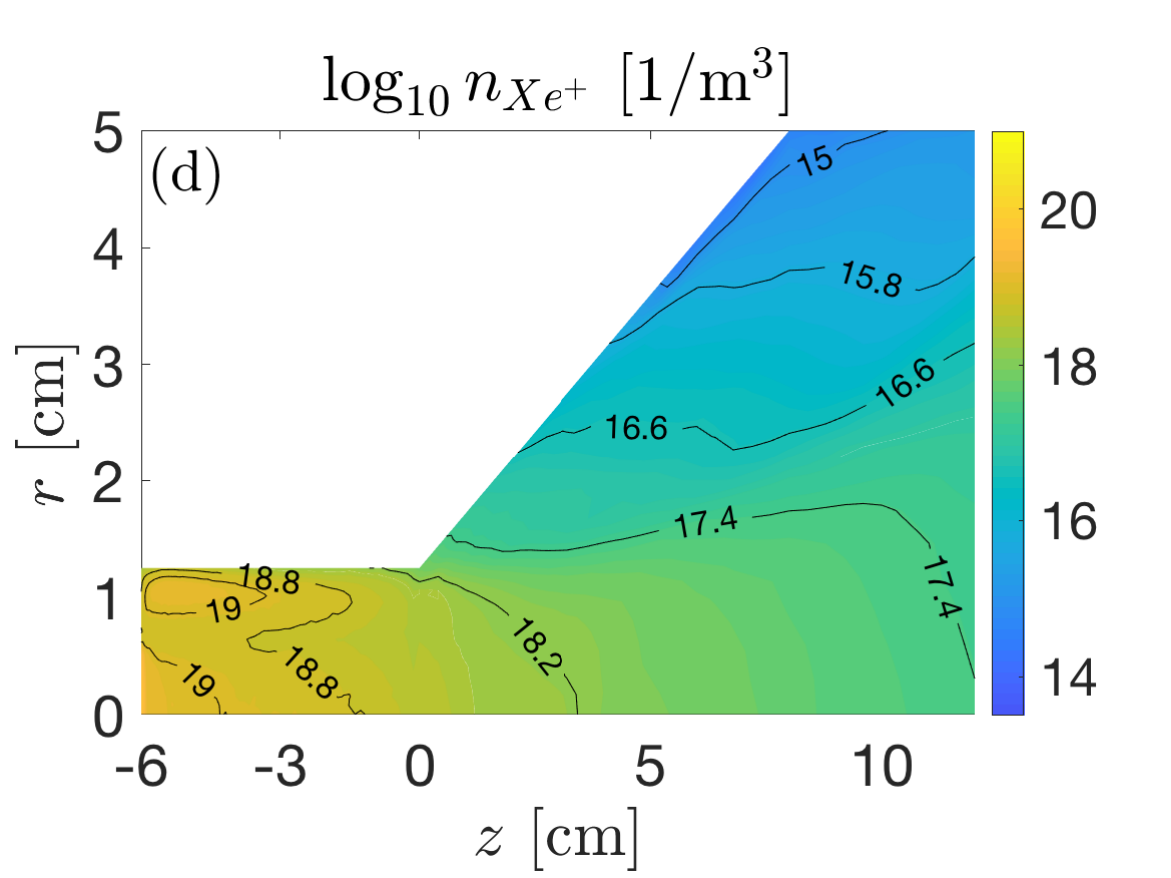}
\end{minipage}

\caption{2D maps of plasma magnitudes for operation with 1mg/s of Xe and 300W (case 1).
The reference potential $\phi=0$ is at the left-bottom point, $(z,r)$[cm]= (-6, 0).
The maximum $\phi$ is 2.51V, at $(z,r)$[cm]= (-5.97, 0).
The maximum $T_e$ is 11.85eV, at $(z,r)$[cm]= (-2.62, 0.62).
}
\label{fig:2Dmaps_Xe}
\end{figure}

Figure \ref{fig:2Dmaps_Xe} plots 2D maps of plasma magnitudes for Xe as propellant and 300W (case 1). The important points to highlight here for comparison with air are the following.
The peak $T_e$, about 12eV, makes ionization very efficient, and explains that the maxima of $\phi$ and $n_e(=n_{Xe^+})$ are very close to the injection. This implies a large recombination at the walls and re-ionization, very well illustrated by the 2D map of $n_{Xe}$. 
Re-ionization makes $n_{Xe^+}$ decay only after the channel exit, which is caused  mainly by ion acceleration. The decay of $n_{Xe}$ is due to geometric expansion instead.
Although $P_a$ peaks at $(z,r)$[cm]=(-3, 0), the peak of $T_e$ does take place 
a bit downwards and at an intermediate radial position. This is a consequence of the radial location of the different contributions to the energy balance: electromagnetic power deposition, inelastic collisions, wall interaction and energy fluxes. The potential drop along the axis (until the end of the simulated plume) is about 50V, which is consistent with an energy of about 35eV based on a specific impulse of $I_{sp}=7.1$ km/s.

\begin{figure}[H]
\centering

\begin{minipage}[c]{0.32\textwidth}
\includegraphics[width=1\textwidth]{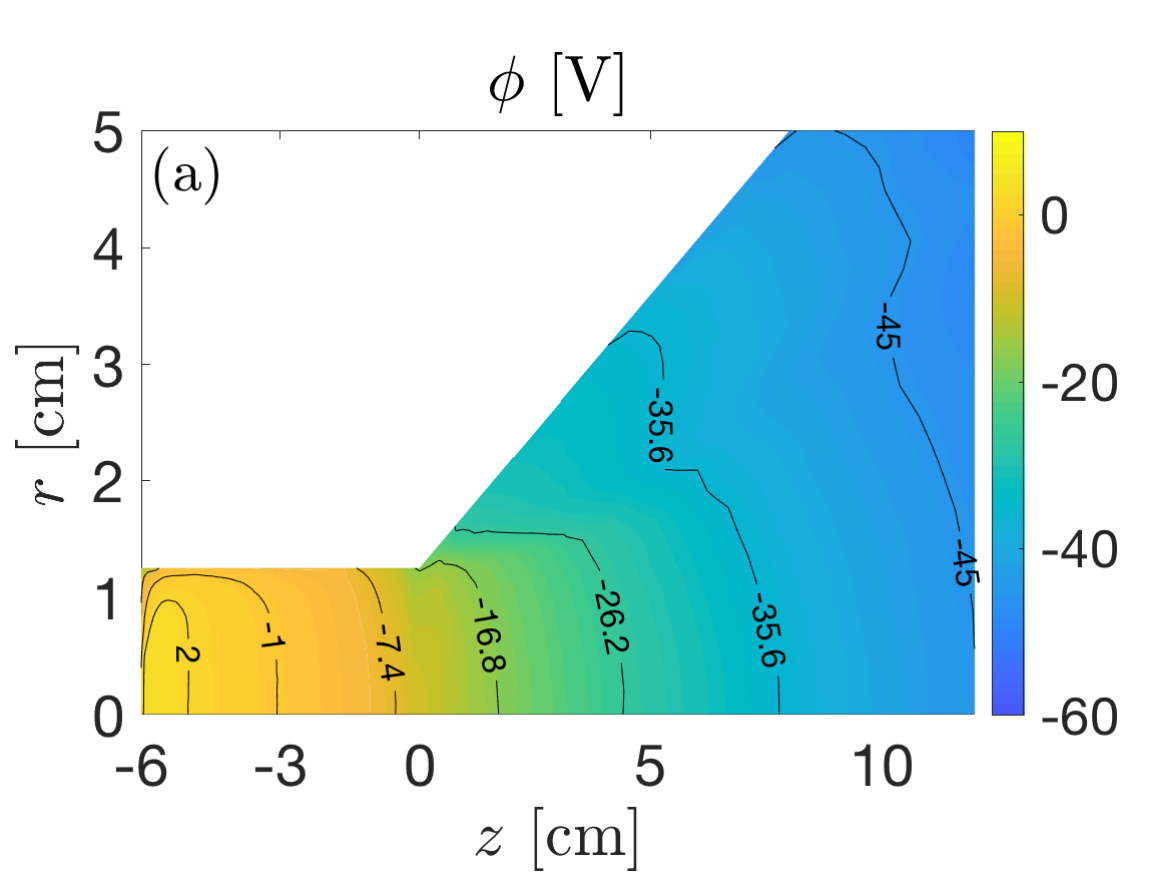}
\end{minipage}
\begin{minipage}[c]{0.32\textwidth}
\includegraphics[width=1\textwidth]{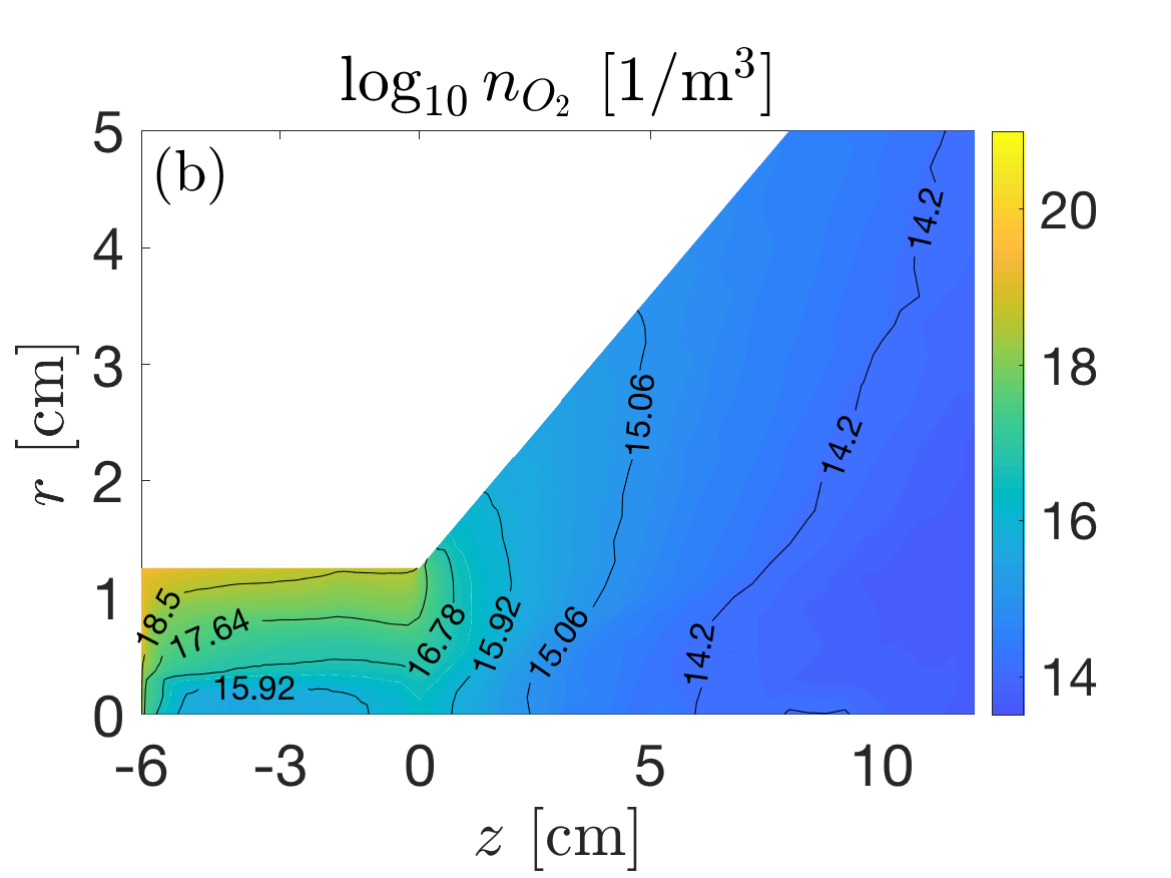}
\end{minipage}
\begin{minipage}[c]{0.32\textwidth}
\includegraphics[width=1\textwidth]{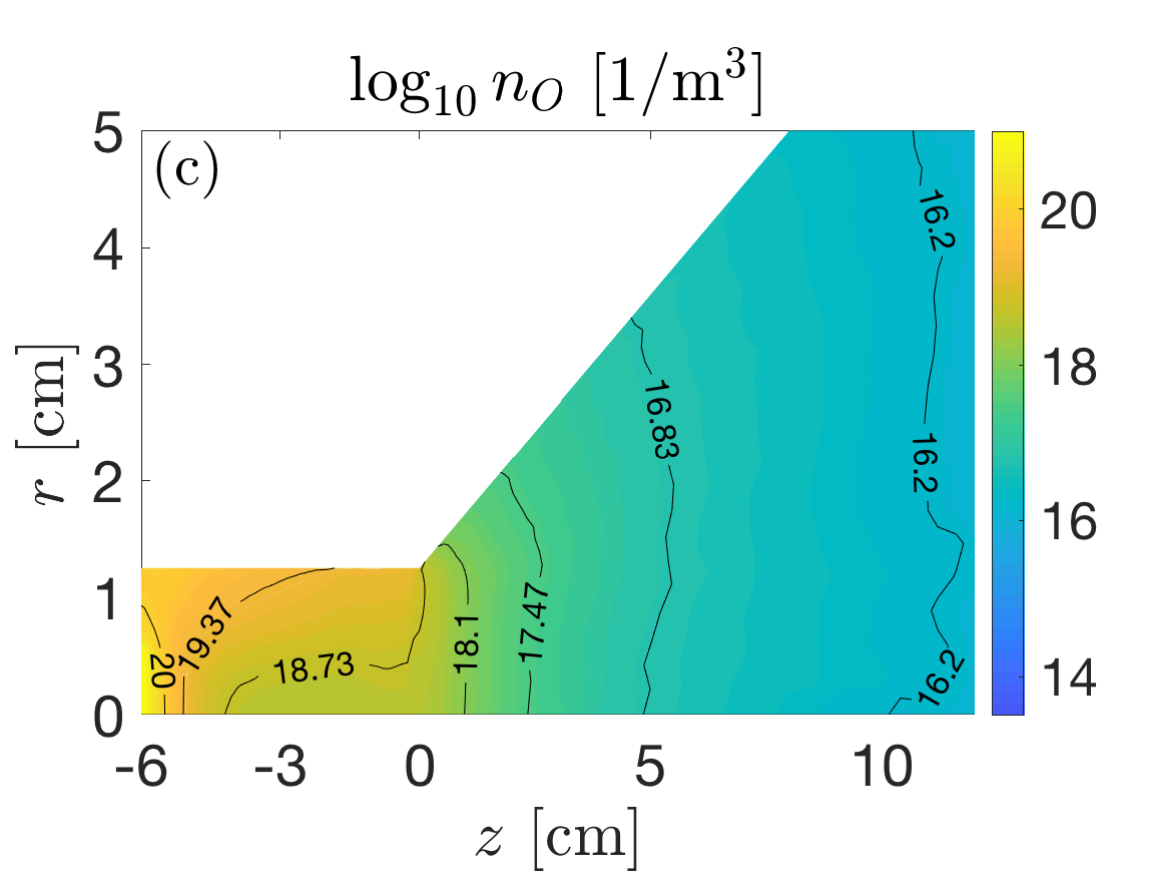}
\end{minipage}

\begin{minipage}[c]{0.32\textwidth}
\includegraphics[width=1\textwidth]{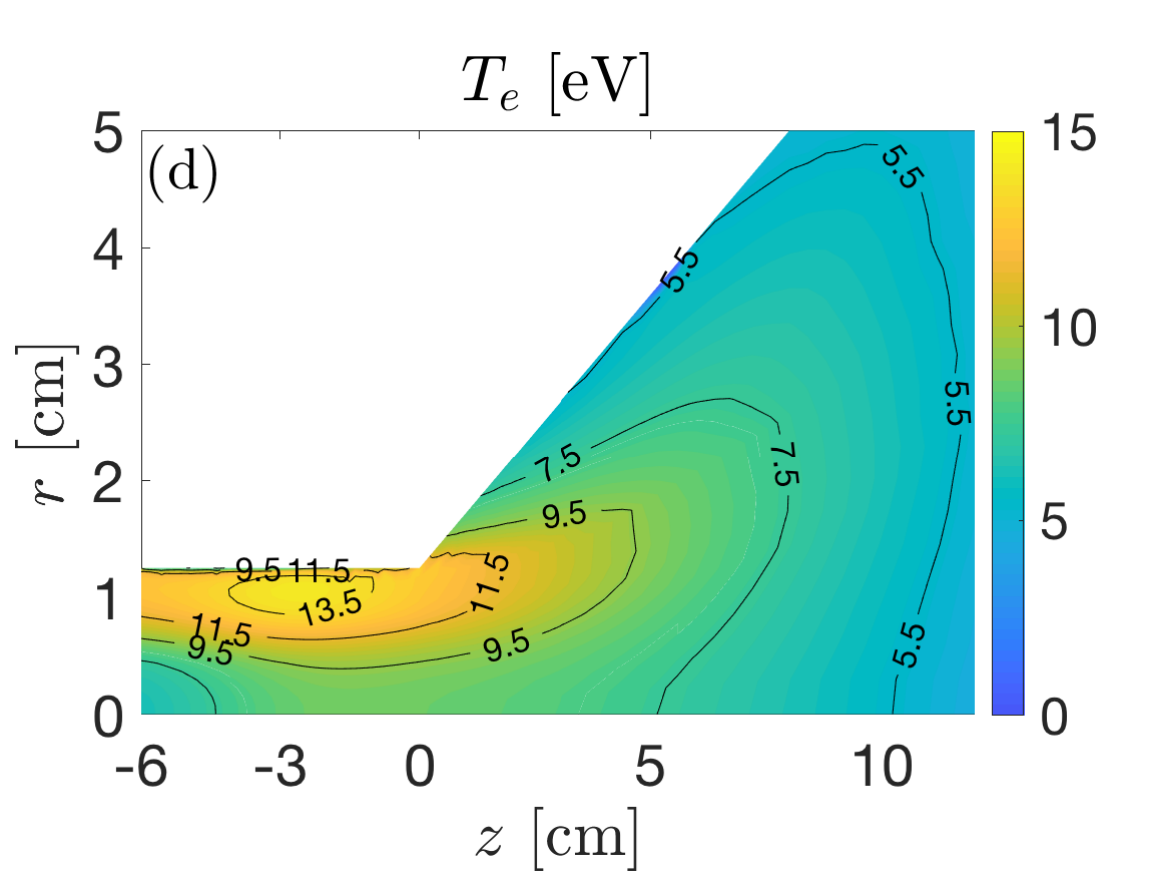}
\end{minipage}
\begin{minipage}[c]{0.32\textwidth}
\includegraphics[width=1\textwidth]{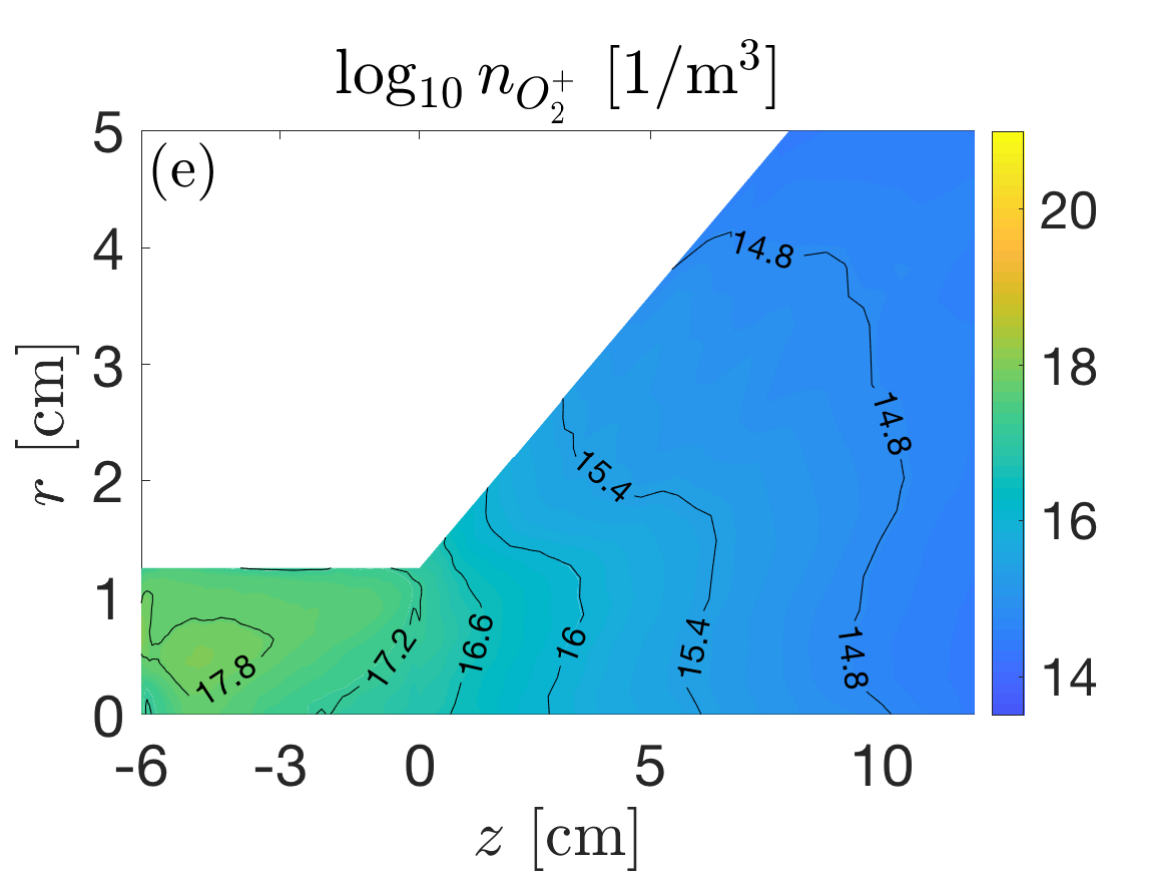}
\end{minipage}
\begin{minipage}[c]{0.32\textwidth}
\includegraphics[width=1\textwidth]{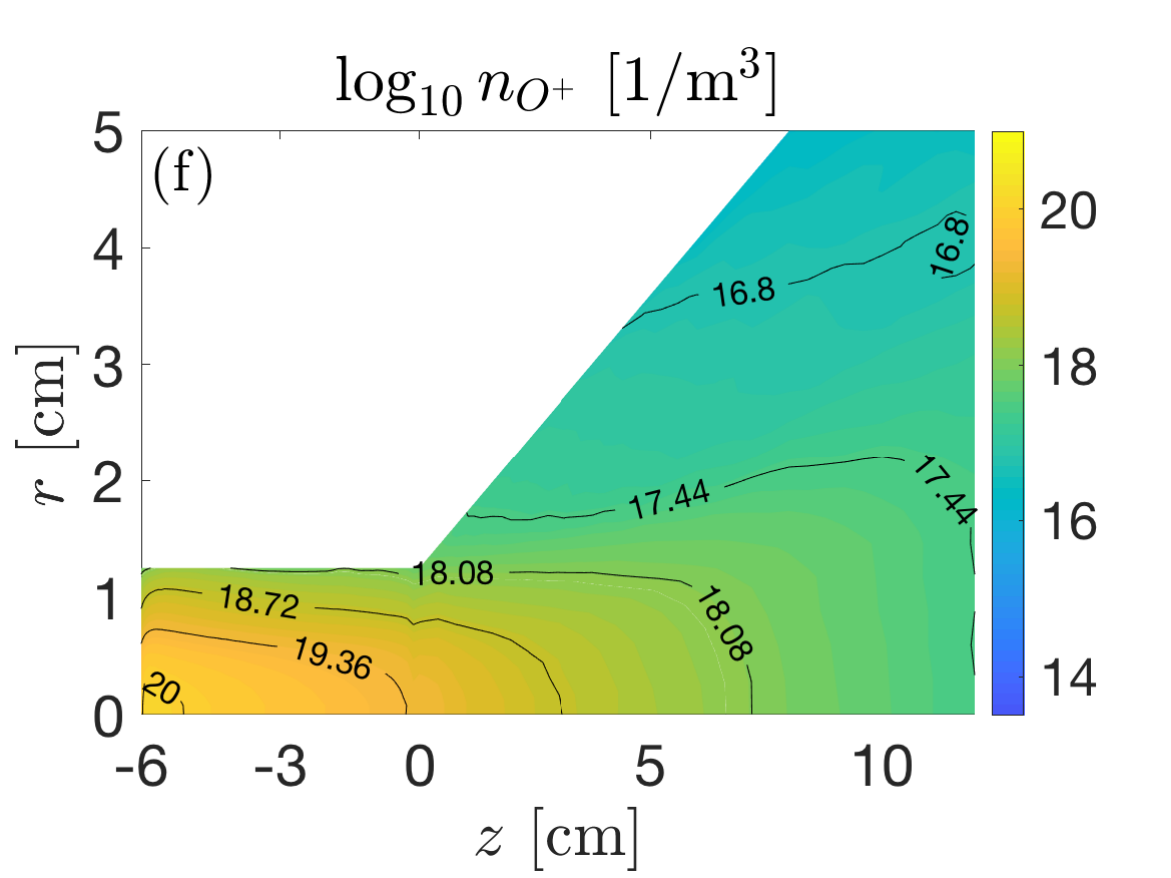}
\end{minipage}

\caption{2D maps of plasma magnitudes for operation with 1mg/s of O and 1500W (case 3).
The reference potential $\phi=0$ is at the left-bottom point, $(z,r)$[cm]= (-6, 0).
The maximum potential is 3.29V,  at $(z,r)$[cm]= (-5.82, 0).
The maximum $T_e$ is 14.17eV, at $(z,r)$[cm]= (-2.63, 1.06).}
\label{fig:2Dmaps_O_1500W}
\end{figure}

Figure \ref{fig:2Dmaps_O_1500W} plots 2D maps for O as propellant and 1500W (case 3). Here the peak $T_e$ is higher, about 14.2eV, and the ionization is efficient.
The maxima of $\phi$ and $n_{O^+}$ are close to the injection, and wall recombination and re-ionization are also high but lower than case 1 (i.e. $\eta_{prod}$ is lower).
The peak of $T_e$ here is also close to the radial wall. The potential drop along the axis is about 45V 
(while 32.6eV is the energy corresponding to O$^+$ based on a specific impulse of 19.8km/s). 
Molecular oxygen is produced by associative recombination at the walls, but its density is marginal compared to the one for atomic oxygen, i.e. $n_O>n_{O_2}$. The same situation is found with the two ion species, $n_{O^+}>n_{O_2^+}$.
Only a 2\% of the ion mass flows downstream corresponds to O$_2^+$.

\begin{figure}[H]
\centering

\begin{minipage}[c]{0.32\textwidth}
\includegraphics[width=1\textwidth]{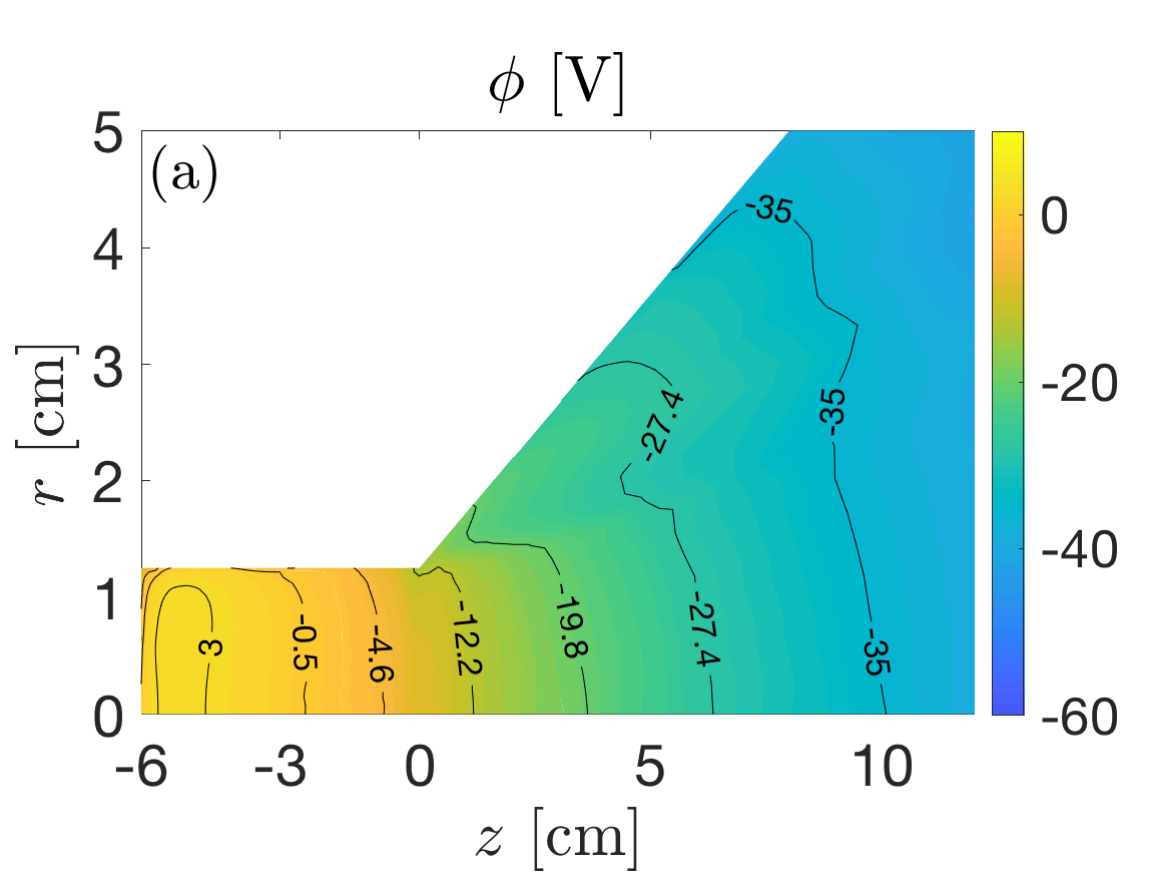}
\end{minipage}
\begin{minipage}[c]{0.32\textwidth}
\includegraphics[width=1\textwidth]{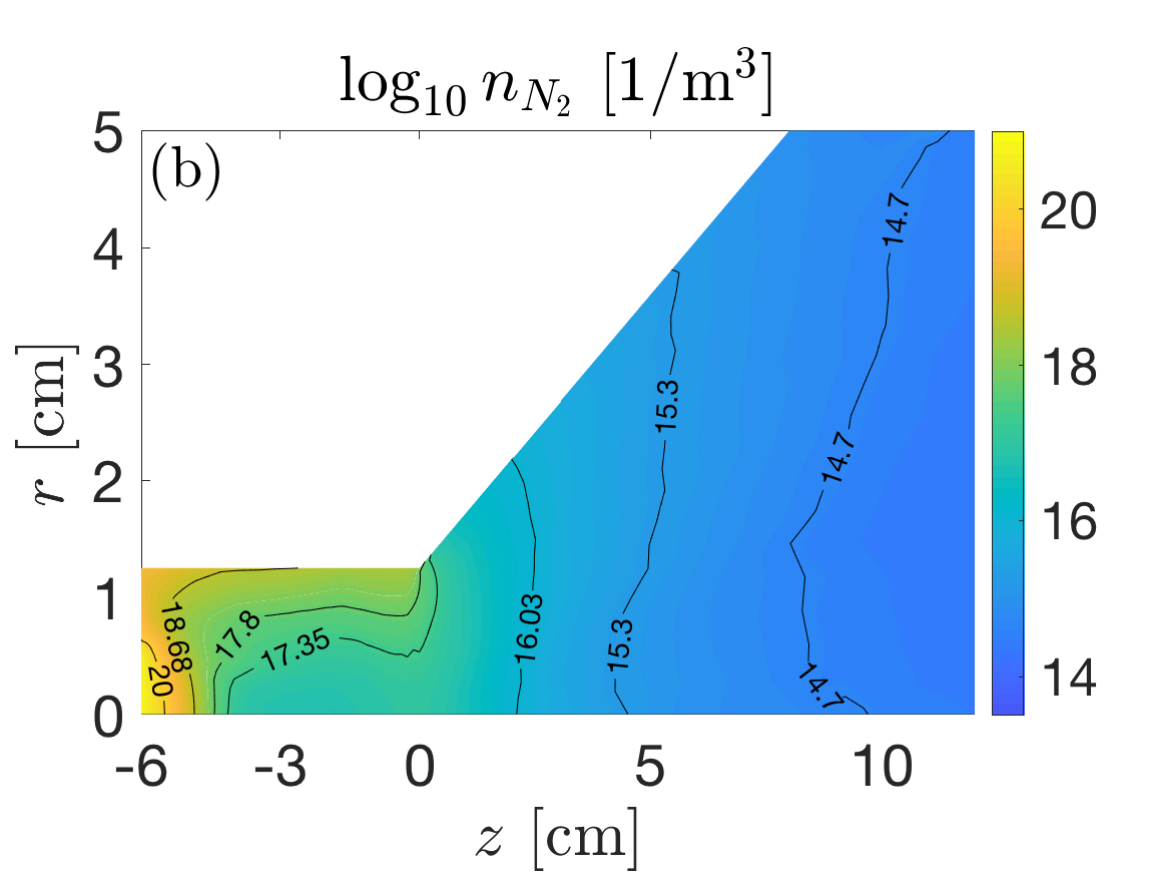}
\end{minipage}
\begin{minipage}[c]{0.32\textwidth}
\includegraphics[width=1\textwidth]{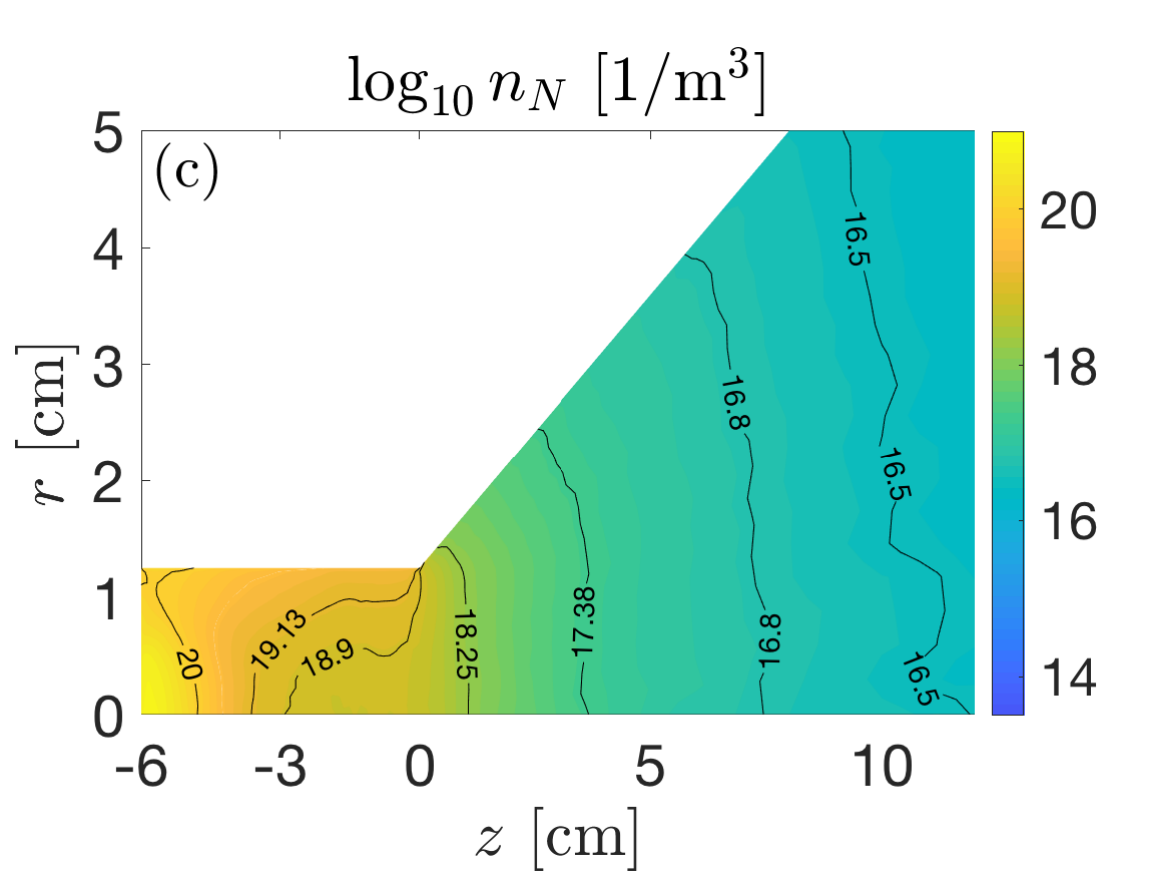}
\end{minipage}

\begin{minipage}[c]{0.32\textwidth}
\includegraphics[width=1\textwidth]{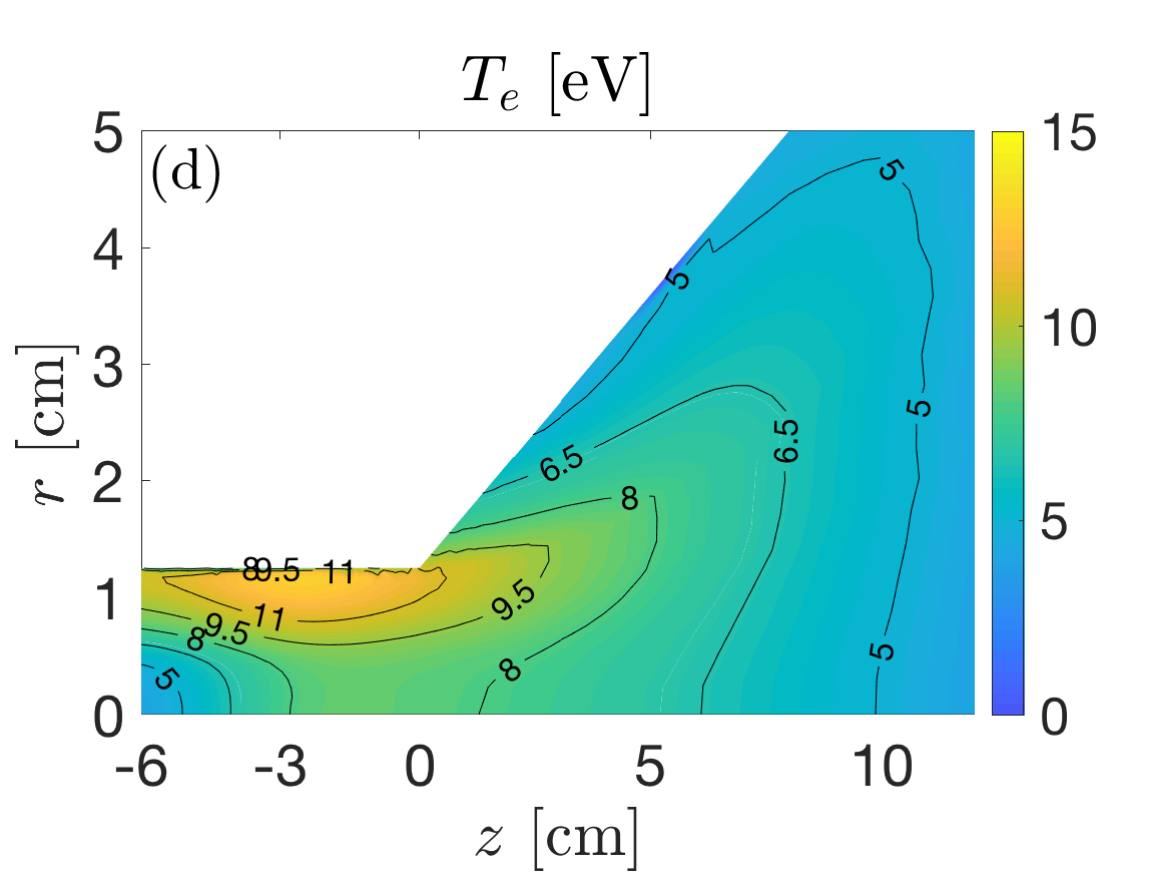}
\end{minipage}
\begin{minipage}[c]{0.32\textwidth}
\includegraphics[width=1\textwidth]{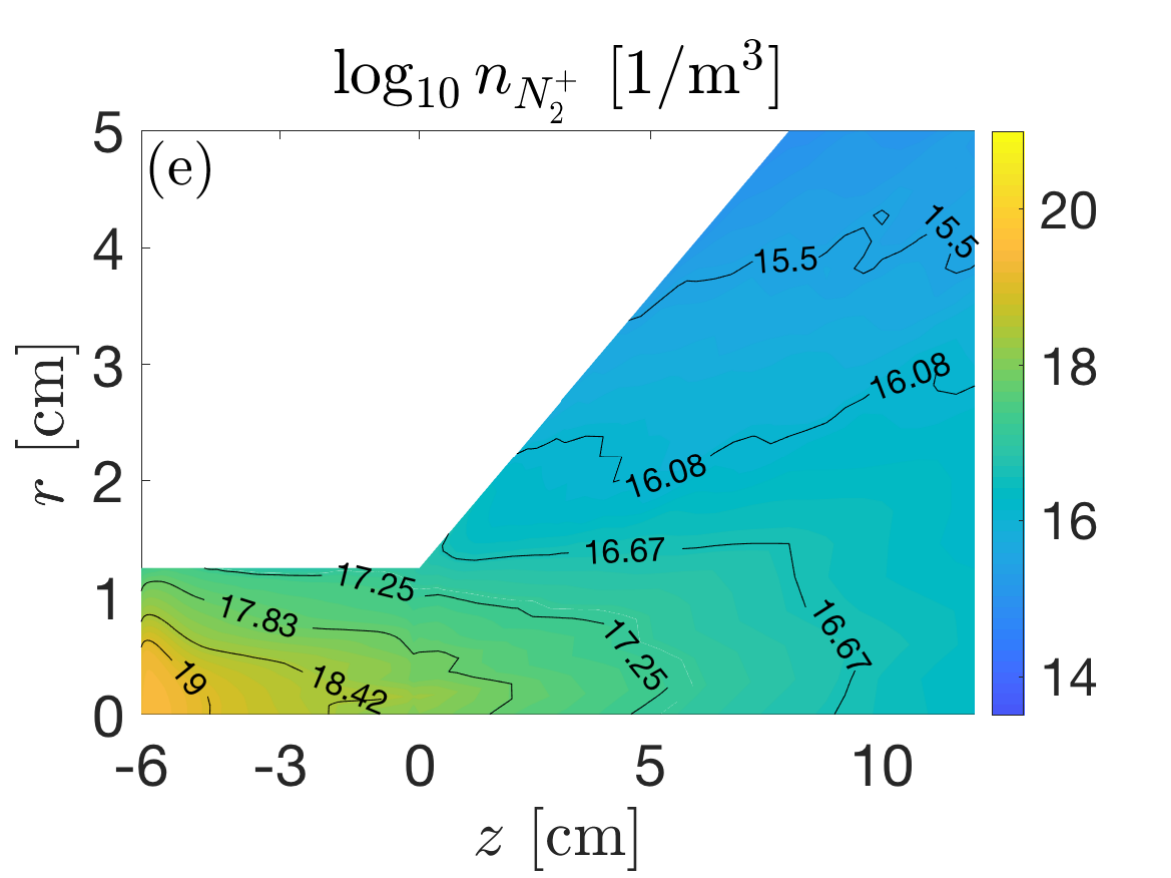}
\end{minipage}
\begin{minipage}[c]{0.32\textwidth}
\includegraphics[width=1\textwidth]{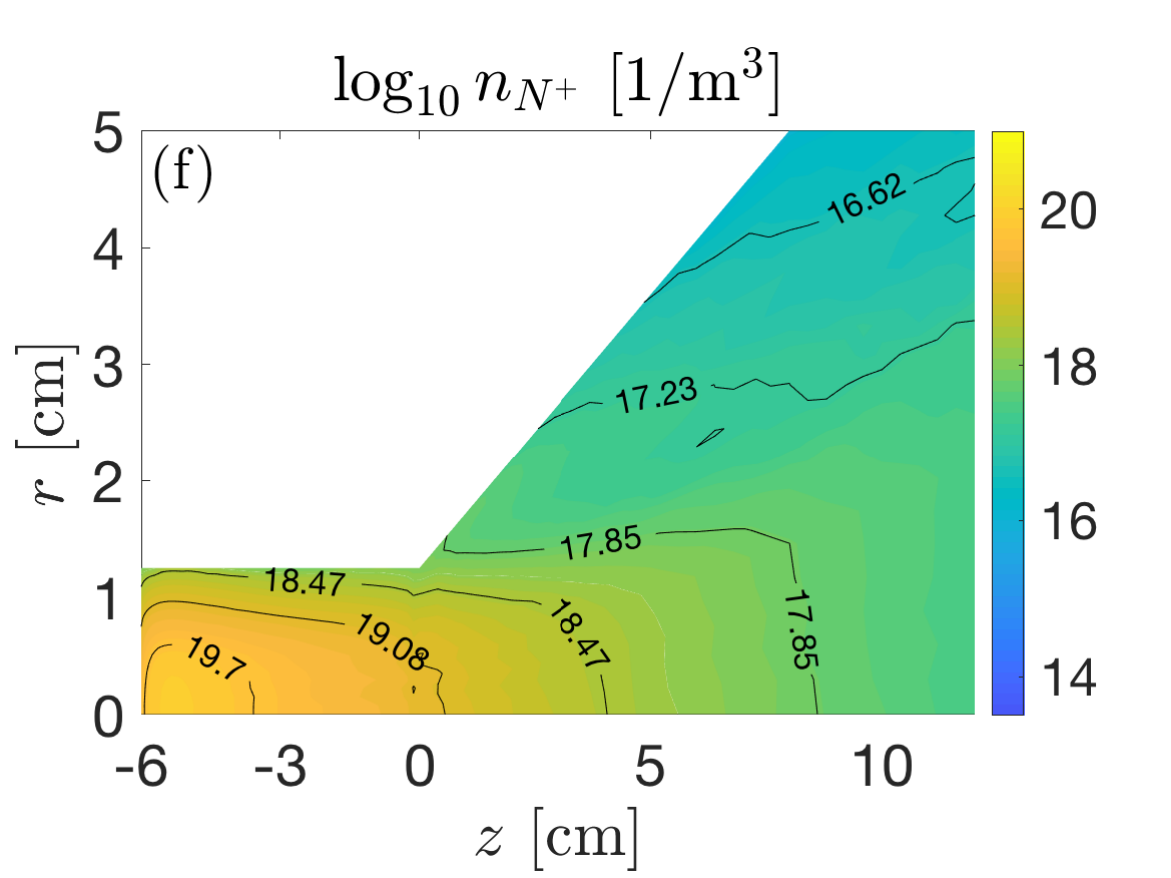}
\end{minipage}

\caption{2D maps of plasma magnitudes 
for operation with 1mg/s of N$_2$ and 2000W (case 6).
The reference potential $\phi=0$ is at the left-bottom point, $(z,r)$[cm]= (-6, 0).
The maximum potential is 3.71V,  at $(z,r)$[cm]= (-5.26, 0).
The maximum $T_e$ is 13.16eV, at $(z,r)$[cm]= (-2.57, 1.17).}
\label{fig:2Dmaps_N2_2000W}
\end{figure}

Figure \ref{fig:2Dmaps_N2_2000W} plots 2D maps for N$_2$ as propellant and 2000W (case 6). The behavior is similar to case 3. The peak of $T_e$ is lower, about 13.2eV. The amount of ionization is lower, the maxima of $\phi$ and $n_N^+$ are shifted slightly downstream, and the amount of wall recombination and re-ionization is slightly lower (i.e. $\eta_{prod}$ is slightly higher). The peak of $T_e$ is also close to the radial wall. The lower $T_e$ explains partially that the potential drop along the axis is about 40V (again consistent with the specific impulse). In terms of plasma composition, there is strong dissociation of N$_2$ just after injection, and then $n_N>n_{N_2}$
in most of the channel and plume. The fraction of atomic ions is much higher than molecular ions, $n_{N^+}>n_{N_2^+}$, practically everywhere. Only a 9\% of the ion mass flows downstream corresponds to N$_2^+$.

The 2D plasma maps for cases 1, 3, and 6 show that the dominant ion species are the atomic ones
(Xe$^+$, O$^+$, N$^+$). 
The Xe$^+$ mass is about 9 times the masses of O$^+$ and N$^+$. 
Since, for the 3 cases, the mass flow is the same and $\eta_u\sim 1$, the electron (i.e. plasma) density is in each case proportional to the ion mass, explaining the differences observed in the 2D maps.
Furthermore, looking at the electron energy equation \eqref{eq:ene}, where all terms are proportional to $n_e$ except for $P'''_a$, one deduces that to reach a similar energetic state (i.e. $T_e$),  
the operation with O and N$_2$ requires one order of magnitude more of absorbed power than the operation with Xe. This explains most of the shift of the performance curves shown in Fig. \ref{fig:vspower} between Xe, O and N$_2$.

\begin{figure}[H]
\centering

\begin{minipage}[c]{0.32\textwidth}
\includegraphics[width=1\textwidth]{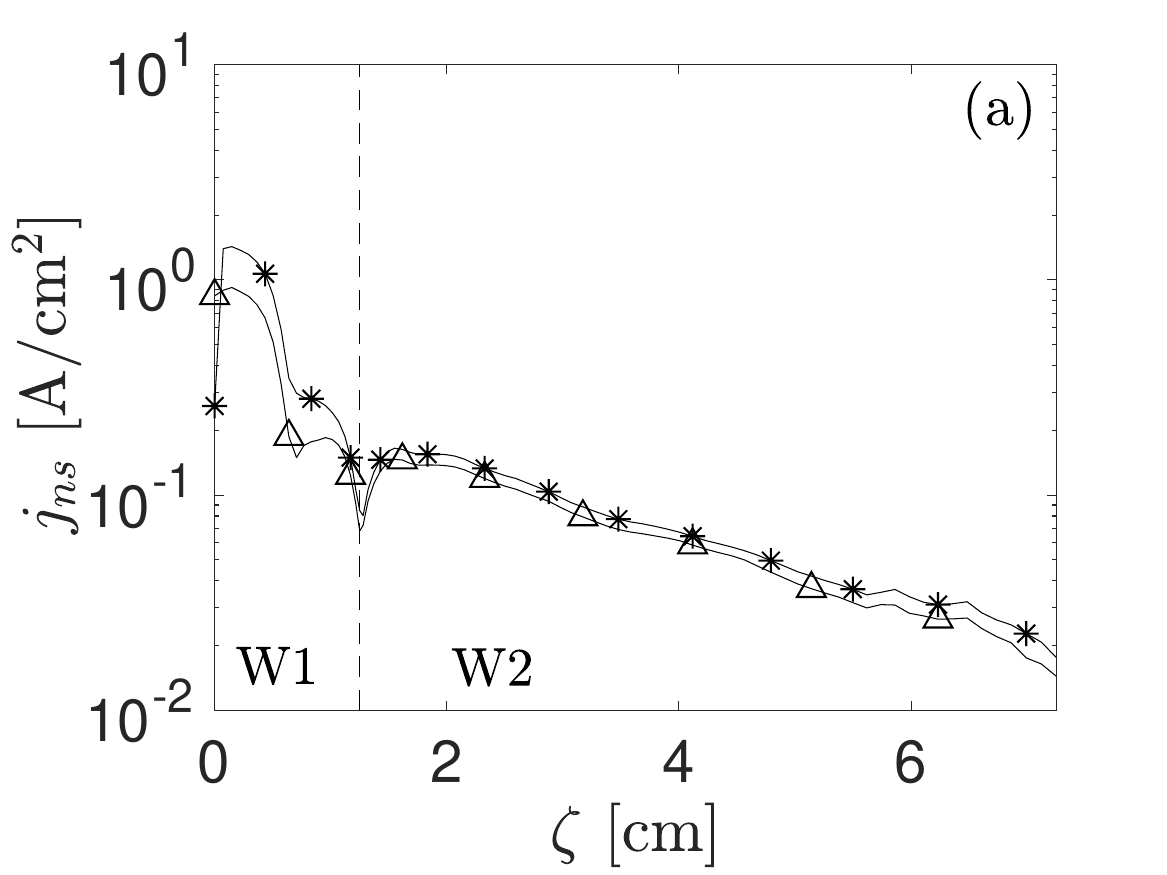}
\end{minipage}
\begin{minipage}[c]{0.32\textwidth}
\includegraphics[width=1\textwidth]{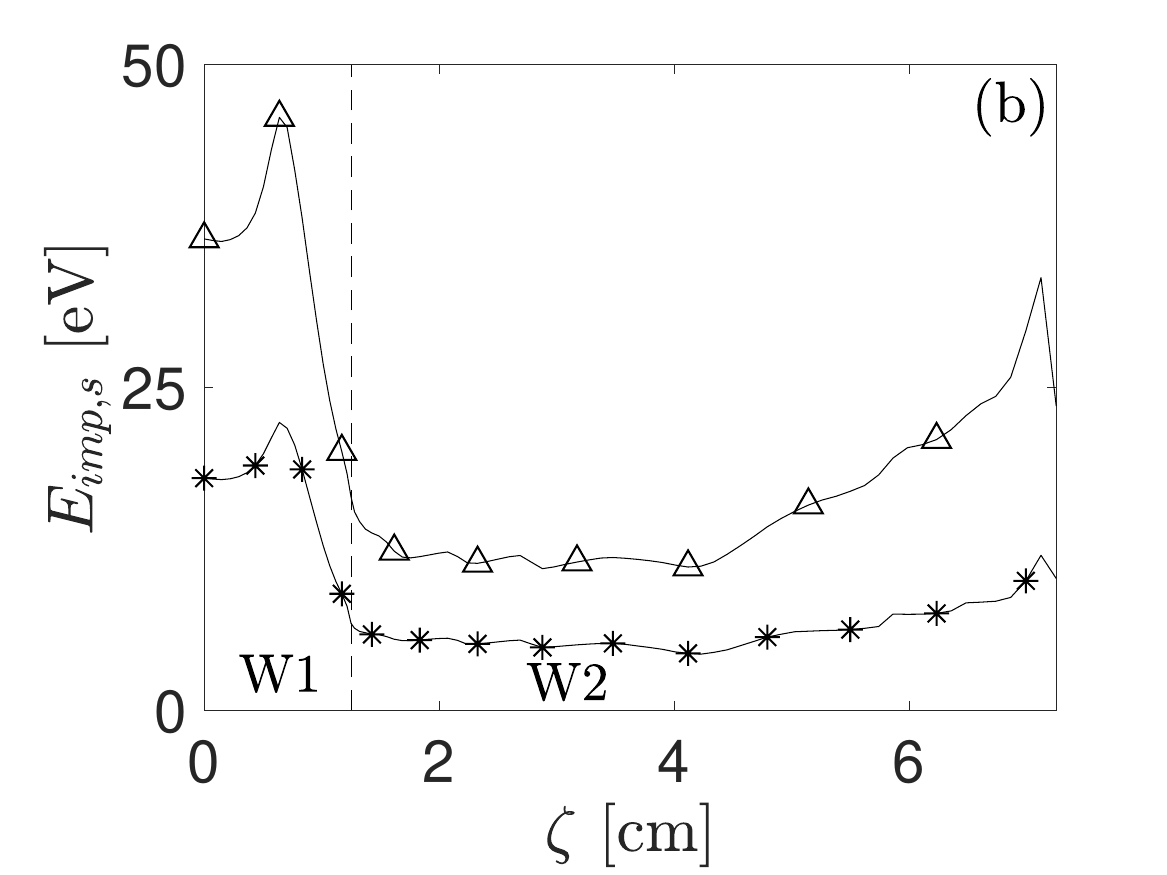}
\end{minipage}
\begin{minipage}[c]{0.32\textwidth}
\includegraphics[width=1\textwidth]{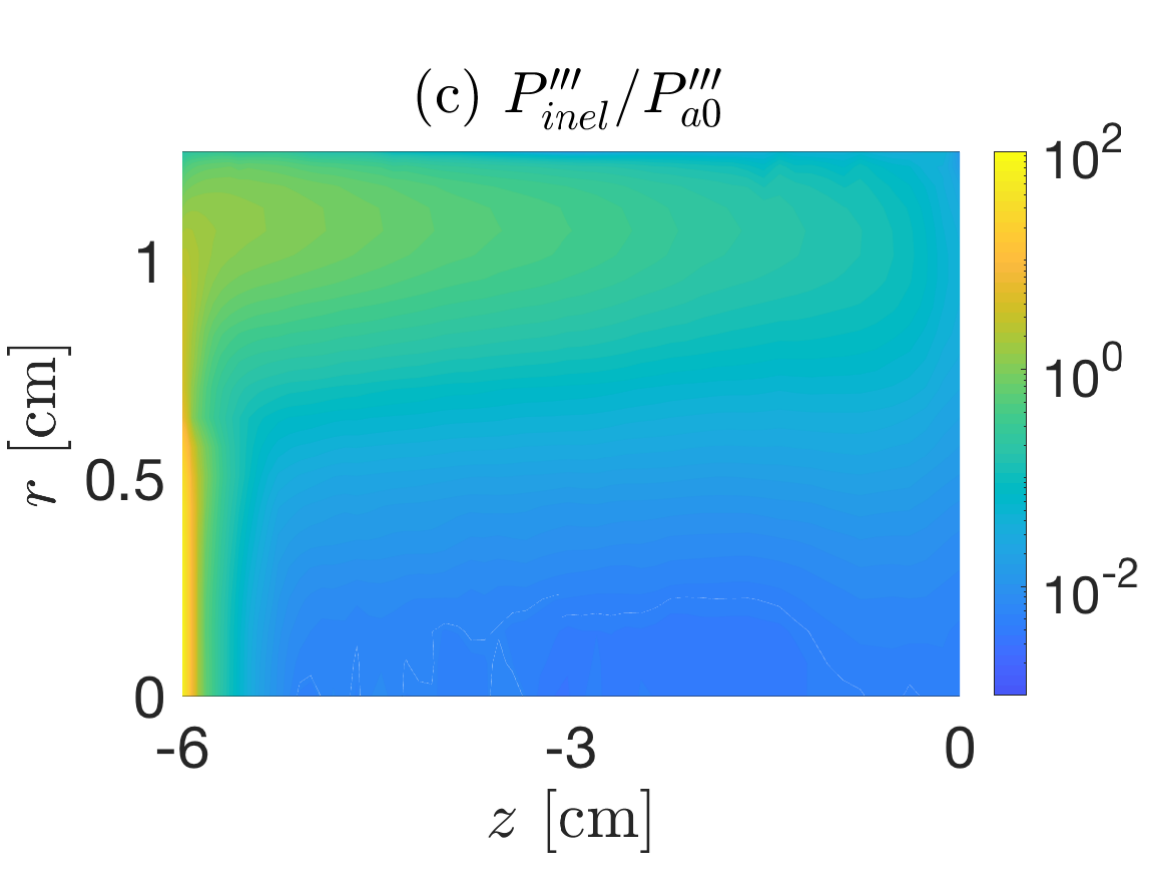}
\end{minipage}

\begin{minipage}[c]{0.32\textwidth}
\includegraphics[width=1\textwidth]{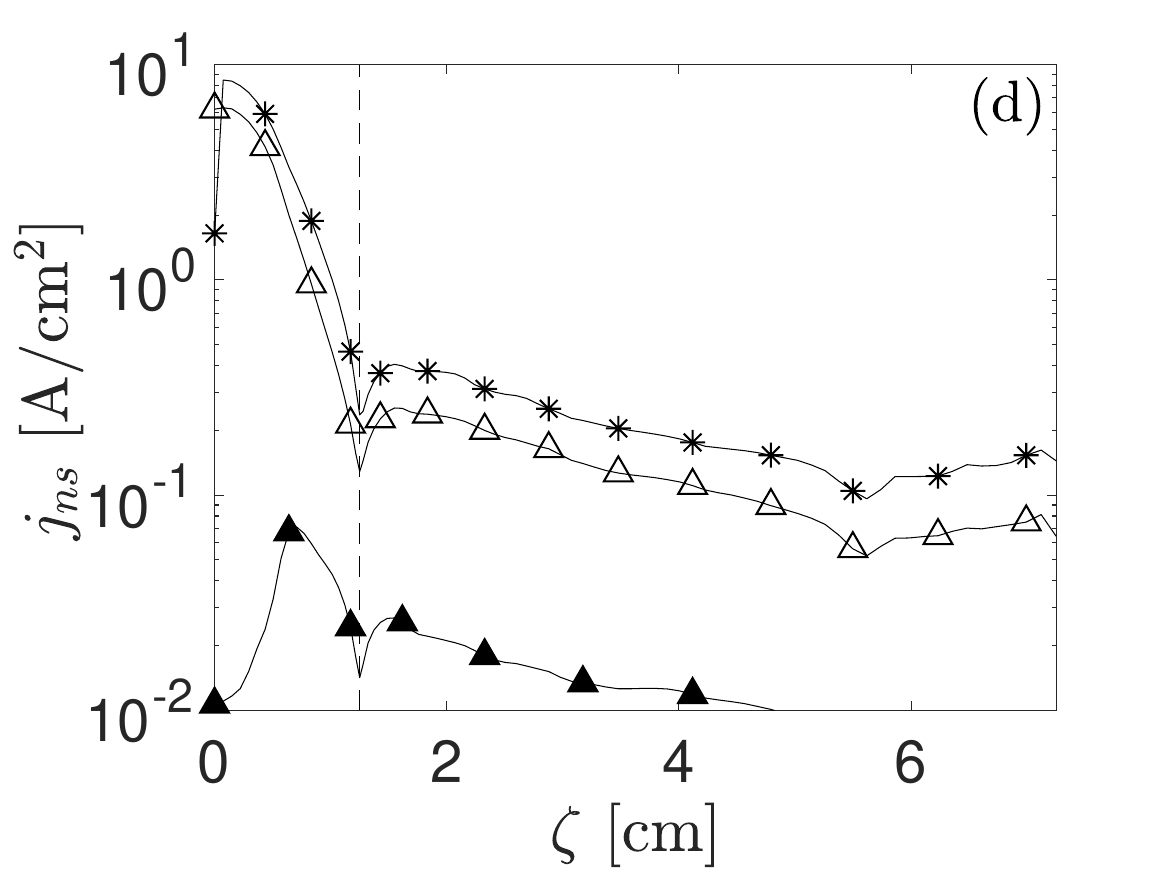}
\end{minipage}
\begin{minipage}[c]{0.32\textwidth}
\includegraphics[width=1\textwidth]{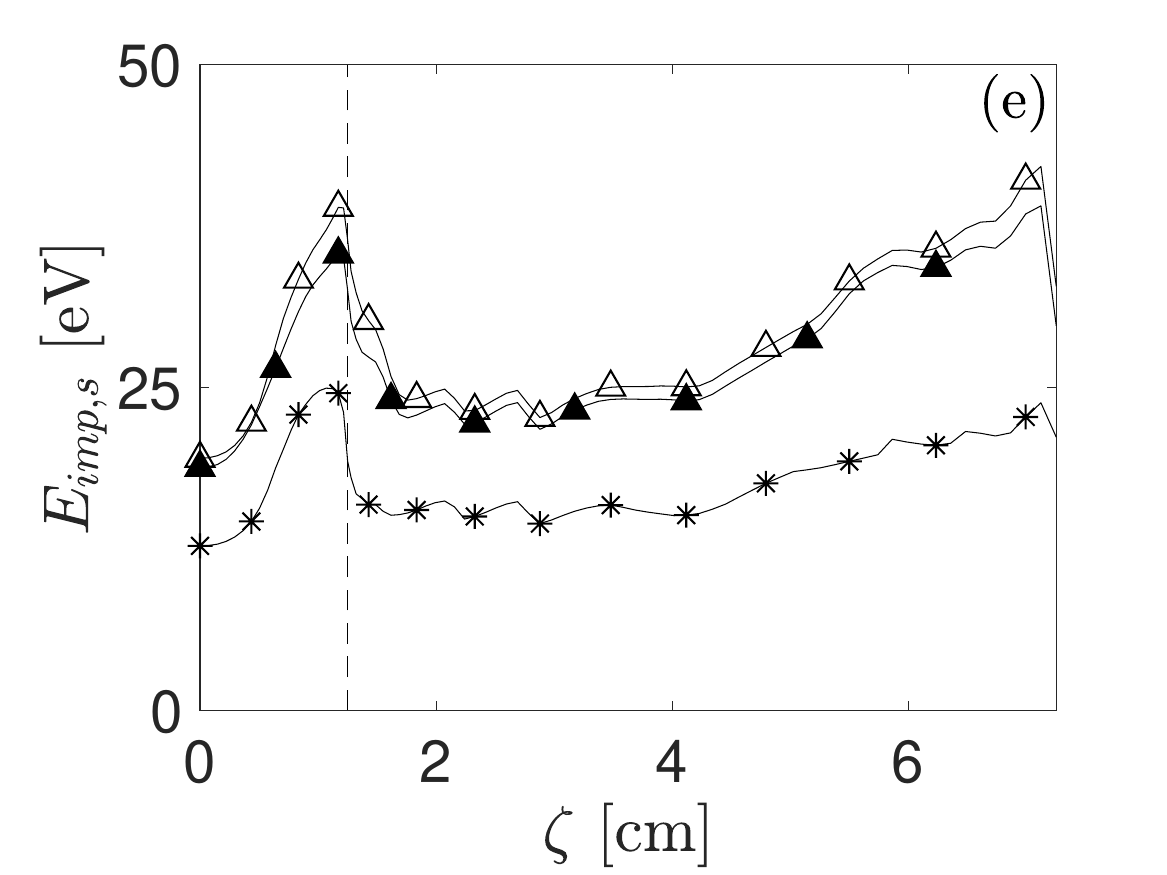}
\end{minipage}
\begin{minipage}[c]{0.32\textwidth}
\includegraphics[width=1\textwidth]{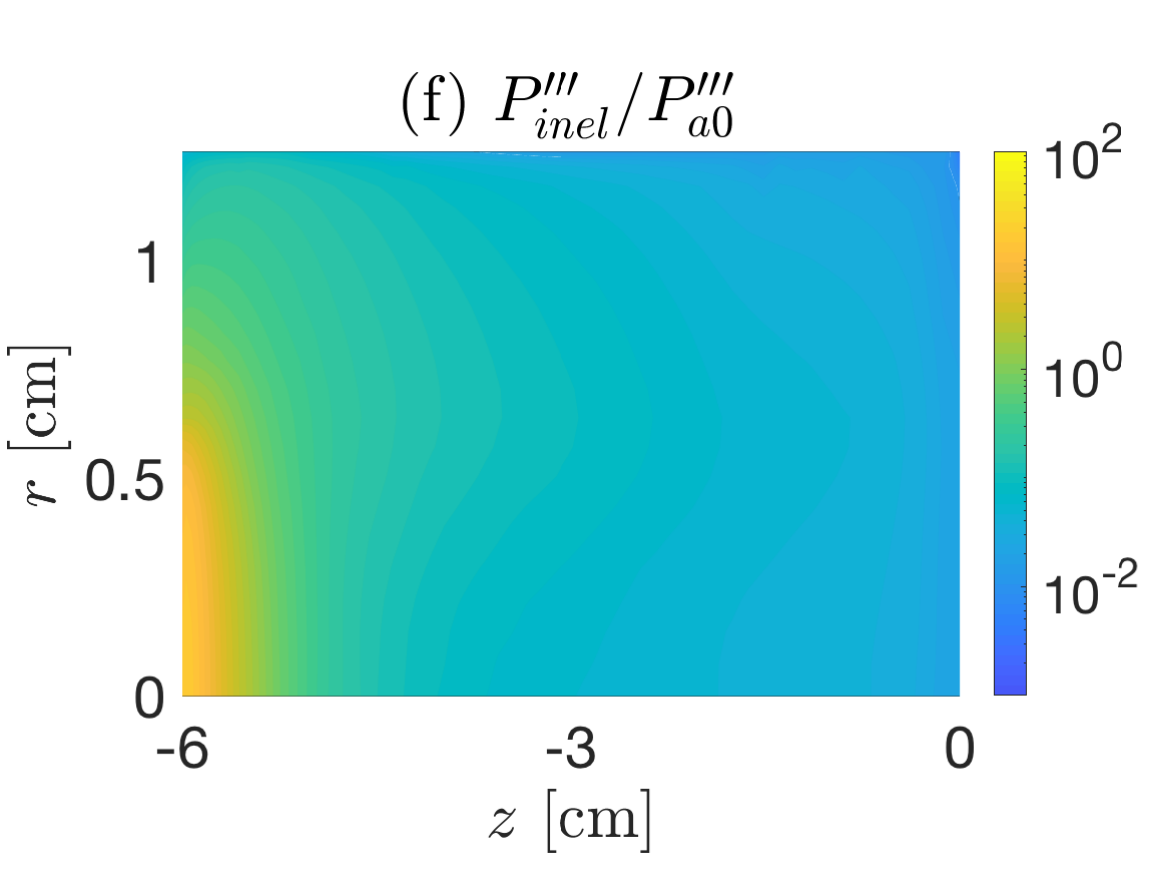}
\end{minipage}

\begin{minipage}[c]{0.32\textwidth}
\includegraphics[width=1\textwidth]{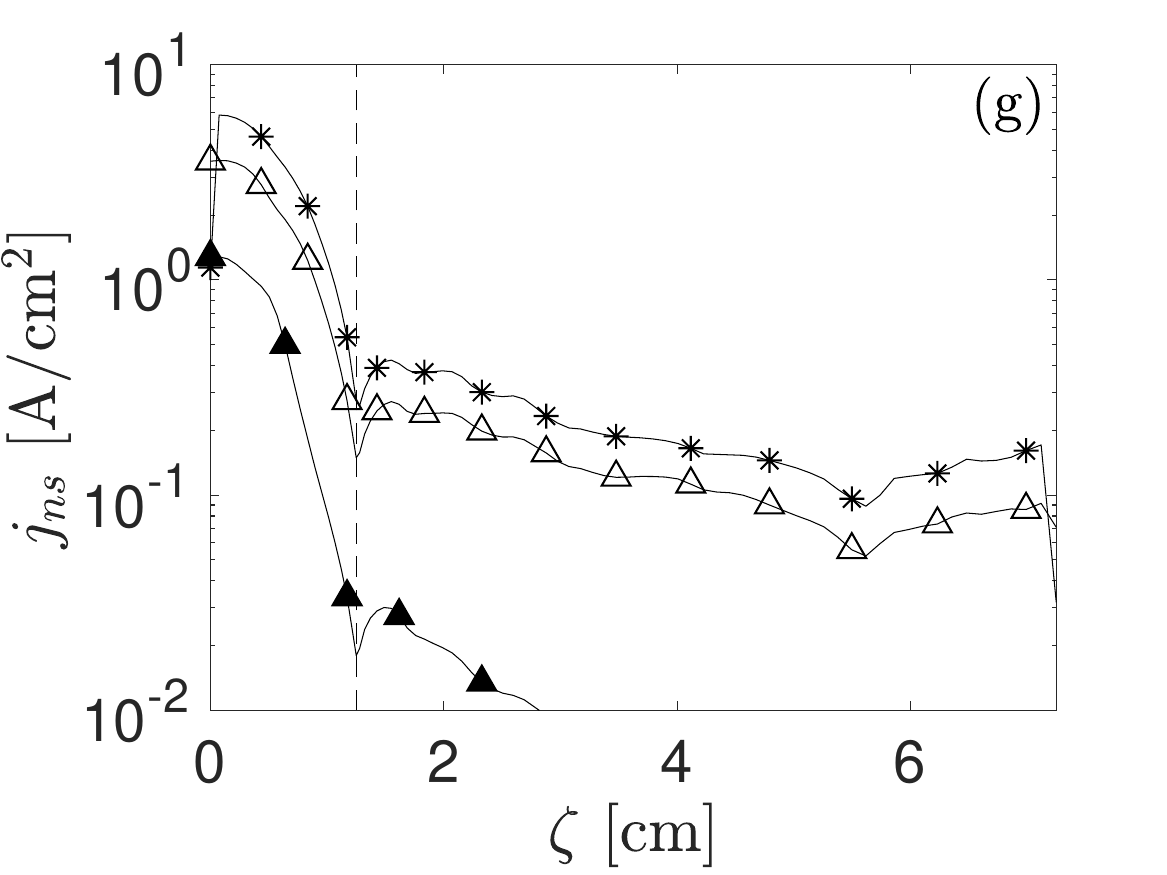}
\end{minipage}
\begin{minipage}[c]{0.32\textwidth}
\includegraphics[width=1\textwidth]{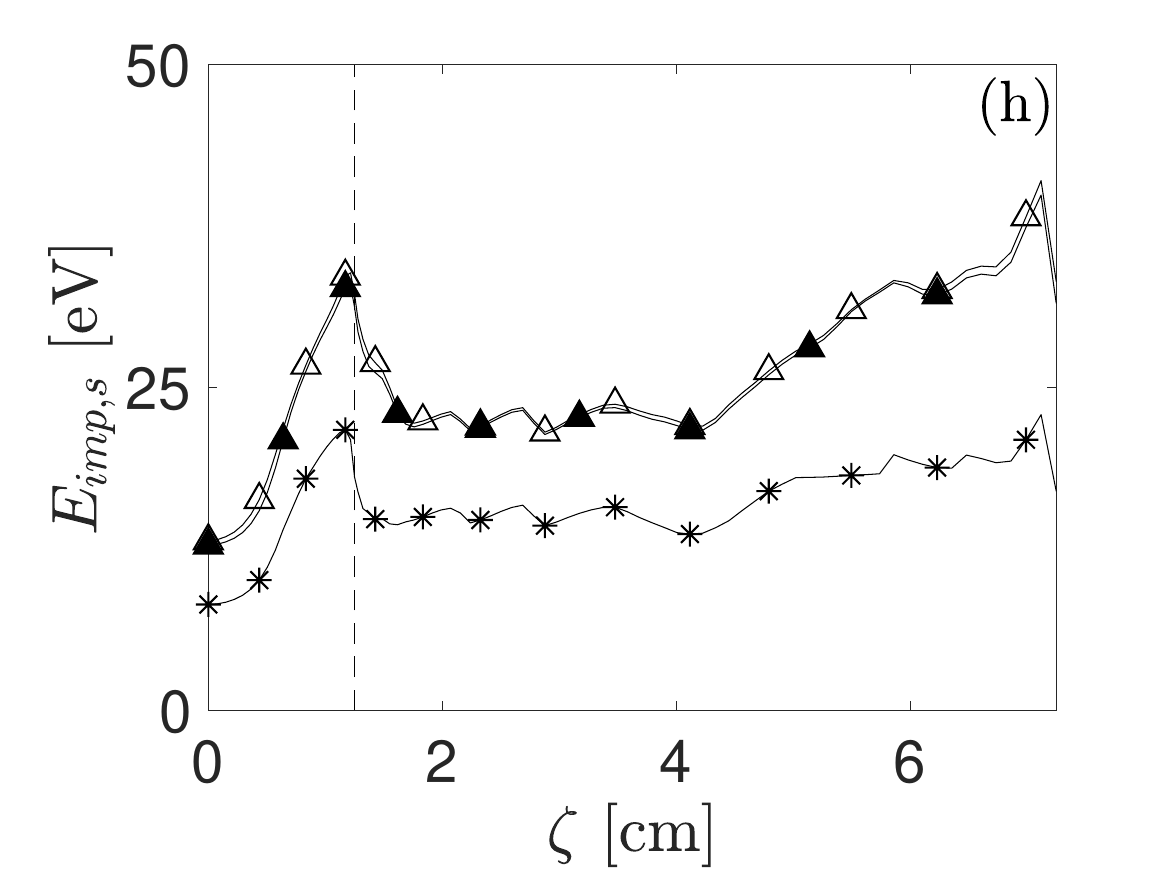}
\end{minipage}
\begin{minipage}[c]{0.32\textwidth}
\includegraphics[width=1\textwidth]{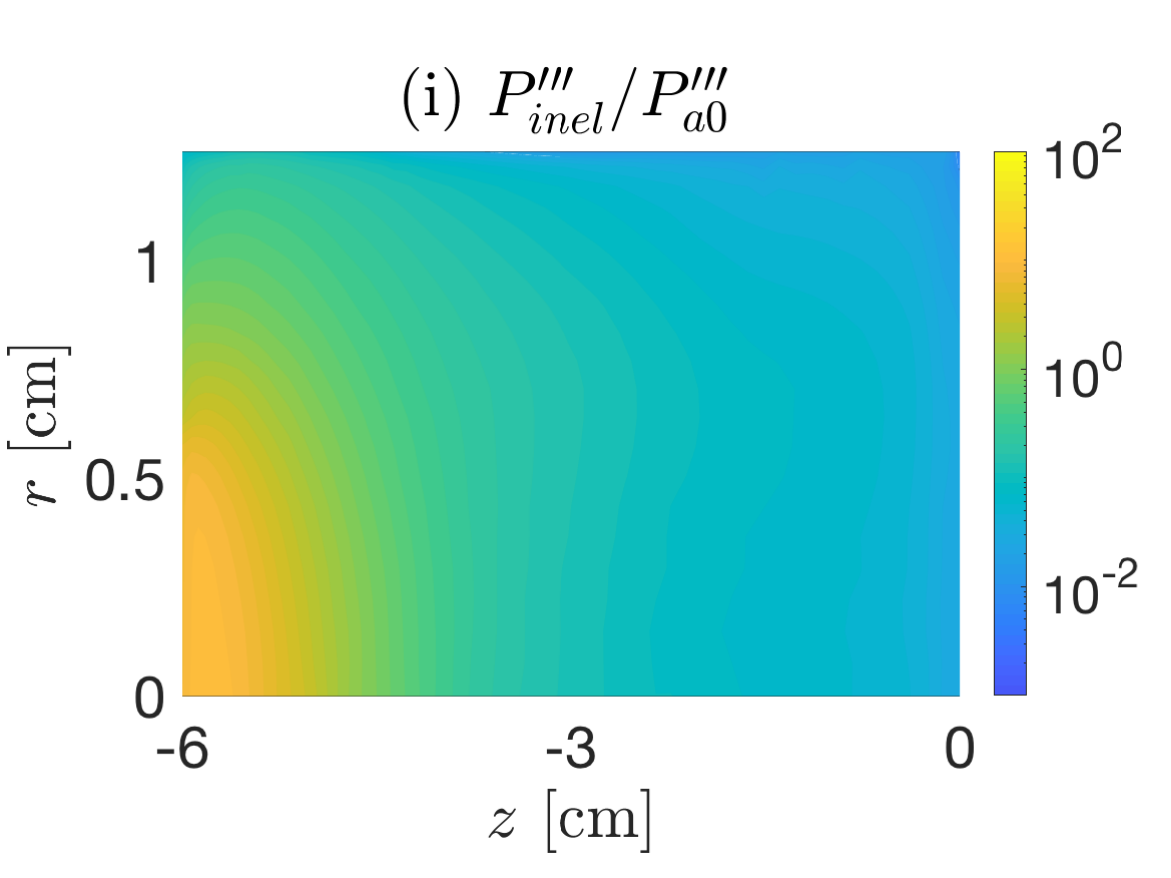}
\end{minipage}

\caption{Particle and energy losses for case 1 (first row, Xe$^+$/$\triangle$, electrons/$\ast$), case 3 (second row, O$_2^+$/$\blacktriangle$, O$^+$/$\triangle$, electrons/$\ast$), and case 6 (third row, N$_2^+$/$\blacktriangle$, N$^+$/$\triangle$, electrons/$\ast$).
First column: current densities to back and lateral walls (W1 and W2);
second column: mean impact energies; and 
third column: inelastic losses, normalized with $P'''_{a0}\equiv P_a/$(source volume). Electron currents and impact energies correspond to the primary electrons from plasma bulk and exclude the SEE.}
\label{fig:powerprof}
\end{figure}

A comparative study of particle and energy losses inside the plasma source for cases 1, 3 and 6 is plotted in Figure \ref{fig:powerprof} (from first to third row, respectively). 
Currents to walls (first column), mean impact energies at walls (second column), and 2D maps of inelastic losses (third column) are shown. 
Current densities and mean impact energies are shown for the charged species, primary electrons and ions, and are plotted along the arc length $\zeta$ of the thruster channel contour, starting at $(z,r)$[cm]$=(-6,0)$ and advancing along the back wall (W1) and the lateral one (W2) as shown in Fig. \ref{fig:thruster_sketch}. 
For each species, 
the mean impact energy is equal to the energy flux divided by the particle flux.
The maps of $P'''_{inel}$ are normalized with $P'''_{a0}$, a constant equal to $P_a$ divided by the source volume.

The response for case 1 (with Xe) is our reference here.
The lack of magnetic confinement in the back wall W1 explains 
that the current densities are larger (by one order of magnitude) there than on the lateral wall W2. Since W1 and W2 are dielectric, the difference between the plotted currents  to the wall
of ions and primary electrons, $j_{ni}-j_{np}$, is  the SEE current from the boron nitride walls.
The mean impact energy of primary electrons is about twice the local $T_e$,
while the mean ion impact energy is about double due to the sheath potential drop. 
The large ion impact energies in W1, in the range 20-45eV, are susceptible to producing erosion \cite{brown20}.
The map of inelastic losses follows that of $n_{Xe}$, which is illustrated by the large ionization and re-ionization near W1 and W2.
Overall, ionization and excitation contribute with 52\% and 48\%, respectively, to $P_{inel}$.

For case 3 (with O), the current to the walls of molecular ions is marginal. For primary electrons and atomic ions, the behavior of the currents to the walls is similar to case 1, except that they are higher due to the higher density for O$^+$.
The impact energies are comparable, in the range 15-25eV for electrons and 20-40eV for ions. The map of inelastic losses also follows mainly that of $n_O$, with inelastic collisions concentrated near the axis instead of the walls due to the associative recombination of O. Ionization and excitation contribute to, respectively, 65\% and 31\%, and dissociation processes only 4\%.

For case 6 (with N$_2$), the behavior is similar to case 3.
Current to the walls, mainly to W1, are slightly lower. Impact energies are lower too due to the lower $T_e$, in the range 10-20eV for electrons and 15-35eV for ions. The inelastic losses are less concentrated around W1 and more around W2. They are distributed as 30\%, 66\% and 4\% for, respectively, ionization, excitation and dissociation processes.

\section{Conclusions}
\label{sec:conc}

The plasma chemistry of the main components of air, nitrogen and oxygen, is implemented in a hybrid simulation code for the plasma discharge in EPTs to assess the capabilities of the air-breathing concept. 
The main collisions between electrons and heavy species for atomic and diatomic substances are considered with elastic momentum transfer, Coulomb interaction, electronic excitation, rotational excitation, vibrational excitation, ionization, and dissociation processes. The main interactions with the thruster walls
(ion recombination and atom associative recombination)
are also considered. The PIC formulation used for heavy species deals with the above collisions
using Monte Carlo collision methods to generate or remove macroparticles of the involved species. The drift-diffusive magnetized fluid model used for electrons accounts for the macroscopic effects of the collisions by including particle and energy sinks and sources in the conservation equations,
and collisional terms in the momentum and heat flux equations. 

The assessment of air-breathing propellants is made by comparing, on a virtual EPT, the performances with Xe and with N$_2$ and O
when operating with a mass flow of 1mg/s and different powers.
The trends of the performances curves are similar but  those of 
N$_2$ and O are shifted to larger powers. 
While the maximum thrust efficiency for Xe is reached  at $\sim$300W, it is reached at $\sim$1250W and $\sim$2000W for O and  N$_2$, respectively. Furthermore, that maximum efficiency is larger for O and N$_2$ than for Xe (13.5\%, 9.9\%, and 8.4\%, respectively) in spite of the additional dissociation and associative recombination losses of the first propellants, due to re-ionization and relatively larger plasma losses (inelastic plus wall losses) for Xe.

Additional cases are run to discuss the propellant mixture. Simulations run injecting a 
50/50 mixture of N$_2$/O, which is close to a realistic composition in low Earth orbit, 
have demonstrated that the plasma response is a 50/50 linear combination of the 
individual propellant responses. This is valid as long as collisions among heavy species are negligible or `identical' for both propellants, and facilitates many parametric studies. 
Other simulations were run injecting O$_2$, since this could be generated at the intake of an air-breathing system due to associative recombination of O. Results show that the plasma response is close to those of O for the operation powers of interest.

From these studies, we point out that the thruster design and operation conditions must be optimized for the types of propellants to be used, 
and it seems to be margin for an efficient-enough plasma thruster working on air.
Codes like HYPHEN are very suitable to carry out the needed parametric studies in the design phase. 

Finally, as future work, we point out that in the first approach for this work, only electron-heavy species collisions were considered. It would be valuable to assess the effects of collisions between heavy species; and the effects of considering state-selective excited states such as metastable states and vibrational states, and the possible stepwise ionization processes from them, which could help in reducing the plasma losses and improving the performances.


\section*{Acknowledgments}
This research was funded initially by the HIPATIA project of HORIZON 2020 (European Commission), Grant No. GA870542, and completed with funding from the SUPERLEO project (Agencia Estatal de Investigaci\'on, Spanish Government), Grant No. TED2021-132484B-I00. 
The stay of J. Zhou  at ISTP-CNR is being supported by the program \textit{Recualificaci\'on del Sistema Universitario Espa\~nol}, \textit{Margarita Salas}, of the Ministerio de Universidades (Spanish Government).

\newpage
\bibliographystyle{unsrturl}
\bibliography{ep2,others}

\begin{thebibliography}{10}

\bibitem{pari10a}
G.~Parissenti, N.~Koch, D.~Pavarin, E.~Ahedo, K.~Katsonis, F.~Scortecci, and M.~Pessana.
\newblock Non conventional propellants for electric propulsion applications.
\newblock In {\em Space Propulsion 2010}, SP2010-1841086, San Sebasti{\'a}n, Spain, 2010.

\bibitem{holste15}
K.~Holste and et~al.
\newblock In search of alternative propellants for ion thrusters.
\newblock In {\em 34th International Electric Propulsion Conference, Hyogo-Kobe, Japan}, IEPC-2015-320, 2015.

\bibitem{gian16}
V.~Giannetti and et~al.
\newblock Electric propulsion system trade-off analysis based on alternative propellant selection.
\newblock In {\em Space Propulsion Conference}, number SP2016-3125194, Rome, Italy, 2016. Association A\'{e}ronautique et Astronautique de France.

\bibitem{vinci21a}
Alfio~E Vinci and St{\'e}phane Mazouffre.
\newblock Direct experimental comparison of krypton and xenon discharge properties in the magnetic nozzle of a helicon plasma source.
\newblock {\em Physics of Plasma}, 28:033504, 2021.

\bibitem{szab13b}
J.~Szabo, M.~Robin, S.~Paintal, B.~Pote, V.~Hruby, and C.~Freeman.
\newblock Iodine propellant space propulsion.
\newblock In {\em 33th International Electric Propulsion Conference, Washington, USA}, IEPC-2013-311, 2013.

\bibitem{bello22}
N.~Bellomo, M.~Magarotto, M.~Manente, F.~Trezzolani, R.~Mantellato, L.~Cappellini, D.~Paulon, A.~Selmo, D.~Scalzi, M.~Minute, M.~Duzzi, A.~Barbato, A.~Schiavon, S.~Di~Fede, N.~Souhair, P.~De~Carlo, F.~Barato, F.~Milza, E.~Toson, and D.~Pavarin.
\newblock Design and in‐orbit demonstration of {REGULUS}, an iodine electric propulsion system.
\newblock {\em CEAS Space Journal}, 14:79--90, 2022.
\newblock \href {https://doi.org/10.1007/s12567-021-00374-4} {\path{doi:10.1007/s12567-021-00374-4}}.

\bibitem{naka18b}
Y.~Nakagawa, H.~Koizumi, H.~Kawahara, and K.~Komurasaki.
\newblock Performance characterization of a miniature microwave discharge ion thruster operated with water.
\newblock {\em Acta Astronautica}, 157:294 -- 299, 2019.

\bibitem{molo19}
R.~Moloney and et~al.
\newblock Experimental validation and performance measurements of an {ECR} thruster operating on multiple propellants.
\newblock In {\em 36th International Electric Propulsion Conference, Vienna, Austria}, IEPC-2019-199, 2019.

\bibitem{seme95}
AV. Semenkin and GO. Chislov.
\newblock Study of anode layer thruster operation with gas mixtures.
\newblock In {\em 24th International Electric Propulsion Conference}, number IEPC-95-78, Moscow, Russia, 1995. Electric Rocket Propulsion Society, Fairview Park, OH.

\bibitem{cara07}
D.~Di~Cara and et~al.
\newblock {RAM} electric propulsion for low {E}arth orbit operation: an {ESA} study.
\newblock In {\em 30th International Electric Propulsion Conference, Florence, Italy}, IEPC-2007-162, 2007.

\bibitem{ande18a}
T.~Andreussi, E.~Ferrato, A.~Piragino, G.~Cifali, A.~Rossodivita, and M.~Andrenucci.
\newblock Development and experimental validation of a {H}all effect thruster {RAM-EP} concept.
\newblock In {\em Space Propulsion Conference}, number SP2018-00431, Seville, Spain, 2018. Association A\'{e}ronautique et Astronautique de France.

\bibitem{hrub22}
V.~Hruby, K.~Hohman, and J.~Szabo.
\newblock Air breathing {H}all effect thruster design studies and experiments.
\newblock In {\em $37^{th}$ International Electric Propulsion Conference}, number IEPC-2022-446, Cambridge, Massachusetts,USA, 2022. Electric Rocket Propulsion Society, Fairview Park, OH.

\bibitem{roma20}
F.~Romano, Y.-A. Chan, G.~Herdrich, C.~Traub, and et~al.
\newblock {RF} helicon-based inductive plasma thruster ({IPT}) design for an atmosphere-breathing electric propulsion system ({ABEP}).
\newblock {\em Acta Astronautica}, 176:476–483, 2020.
\newblock \href {https://doi.org/10.1016/j.actaastro.2020.07.008} {\path{doi:10.1016/j.actaastro.2020.07.008}}.

\bibitem{ande22b}
T.~Andreussi, E.~Ferrato, and V.~Giannetti.
\newblock A review of air-breathing electric propulsion: from mission studies to technology verification.
\newblock {\em Journal of Electric Propulsion}, 1:31, 2022.

\bibitem{garr12}
L.~Garrigues.
\newblock Computational study of {H}all-effect thruster with ambient atmospheric gas as propellant.
\newblock {\em Journal of Propulsion and Power}, 28(2):344--354, 2012.

\bibitem{tap21}
A.~Taploo, L.~Lin, and M.~Keidar.
\newblock Analysis of ionization in air-breathing plasma thruster.
\newblock {\em Physics of Plasmas}, 28:093505, 2021.

\bibitem{ferr22}
E.~Ferrato, V.~Giannetti, F.~Califano, and T.~Andreussi.
\newblock Atmospheric propellant fed {H}all thruster discharges: 0{D}-hybrid model and experimental results.
\newblock {\em Plasma Sources Science and Technology}, 31(7):075003, 2022.

\bibitem{souh23}
N~Souhair, M~Magarotto, R~Andriulli, and F~Ponti.
\newblock Prediction of the propulsive performance of an atmosphere-breathing electric propulsion system on cathode-less plasma thruster.
\newblock {\em Aerospace}, 10(100), 2023.

\bibitem{tacc22}
F~Taccogna, F~Cichocki, and P~Minelli.
\newblock Coupling plasma physics and chemistry in {PIC} model of electric propulsion: application to an air-breathing low power {H}all thruster.
\newblock {\em Frontiers}, 10:1006994, 2022.
\newblock \href {https://doi.org/10.3389/fphy.2022.1006994} {\path{doi:10.3389/fphy.2022.1006994}}.

\bibitem{roma21a}
F.~Romano and et~al.
\newblock Intake design for an atmosphere-breathing electric propulsion system.
\newblock {\em Acta Astronautica}, 187:225--235, 2021.

\bibitem{roma21b}
F.~Romano and et~al.
\newblock Design, set-up, and first ignition of the rf helicon-based plasma thruster.
\newblock In {\em Space Propulsion Conference 2020+1}, SP2020+1-00247. Association A\'{e}ronautique et Astronautique de France, Paris, France, 2021.

\bibitem{zhou19a}
Jiewei Zhou, Daniel P\'erez-Grande, Pablo Fajardo, and Eduardo Ahedo.
\newblock Numerical treatment of a magnetized electron fluid within an electromagnetic plasma thruster code.
\newblock {\em Plasma Sources Science and Technology}, 28(11):115004, 2019.

\bibitem{pera22b}
Jes{\'{u}}s Perales-D{\'{i}}az, Adri{\'{a}}n Dom{\'{i}}nguez-V{\'{a}}zquez, Pablo Fajardo, Eduardo Ahedo, Farbod Faraji, Maryam Reza, and Tommaso Andreussi.
\newblock Hybrid plasma simulations of a magnetically shielded {H}all thruster.
\newblock {\em Journal of Applied Physics}, 131(10):103302, 2022.
\newblock \href {https://doi.org/10.1063/5.0065220} {\path{doi:10.1063/5.0065220}}.

\bibitem{svil21a}
A.~S\'anchez-Villar, J.~Zhou, M.~Merino, and E.~Ahedo.
\newblock Coupled plasma transport and electromagnetic wave simulation of an {ECR} thruster.
\newblock {\em Plasma Sources Science and Technology}, 30(4):045005, 2021.
\newblock \href {https://doi.org/10.1088/1361-6595/abde20} {\path{doi:10.1088/1361-6595/abde20}}.

\bibitem{zhou22a}
J.~Zhou, A.~Dom\'inguez-V\'azquez, P.~Fajardo, and E.~Ahedo.
\newblock Magnetized fluid electron model within a two-dimensional hybrid simulation code for electrodeless plasma thrusters.
\newblock {\em Plasma Sources Science and Technology}, 31(4):045021, 2022.

\bibitem{tabata06}
T.~Tabata, T.~Shirai, M.~Sataka, and H.~Kubo.
\newblock Analytic cross sections for electron impact collisions with nitrogen molecules.
\newblock {\em Atomic Data and Nuclear Data Tables}, 92:375--406, 2006.

\bibitem{tabata12}
T.~Tabata, T.~Shirai, M.~Sataka, and H.~Kubo.
\newblock Erratum to ‘‘analytic cross sections for electron impact collisions with nitrogen molecules’’ [at. data nucl. data tables 92 (2006) 375-406].
\newblock {\em Atomic Data and Nuclear Data Tables}, 98:74, 2012.

\bibitem{trinitiLXCAT}
Triniti database.
\newblock \url{https://nl.lxcat.net}.
\newblock Accessed: 2019-10-23.

\bibitem{bsrLXCAT}
Bsr database.
\newblock \url{https://nl.lxcat.net}.
\newblock Accessed: 2019-10-23.

\bibitem{istLXCAT}
Ist database.
\newblock \url{https://nl.lxcat.net}.
\newblock Accessed: 2019-10-23.

\bibitem{itik09}
Y.~Itikawa.
\newblock Cross sections for electron collisions with oxygen molecules.
\newblock {\em Journal of Physical and Chemical Reference Data}, 38:1--20, 2009.

\bibitem{biagiLXCAT}
Biagi database.
\newblock \url{https://nl.lxcat.net}.
\newblock Accessed: 2019-10-23.

\bibitem{lawton78}
S.~A. Lawton and A.~V. Phelps.
\newblock Excitation of the $\mathrm{b^1\Sigma_g^+}$ state of o$_2$ by low energy electrons.
\newblock {\em Journal of Chemical Physics}, 69:1055--1068, 1978.

\bibitem{laher90}
R.R. Laher and F.R. Gilmore.
\newblock Updated excitation and ionization cross sections for electron impact on atomic oxygen.
\newblock {\em Journal of Physical and Chemical Reference Data}, 19:277--305, 1990.

\bibitem{haya03}
M.~Hayashi.
\newblock Bibliography of electron and photon cross sections with atoms and molecules published in the 20th century. {X}enon.
\newblock Technical report, National Institute for Fusion Science, 2003.

\bibitem{hayashiLXCAT}
Hayashi database.
\newblock \url{https://nl.lxcat.net}.
\newblock Accessed: 2019-10-23.

\bibitem{taka19}
K.~Takahashi.
\newblock Helicon--type radiofrequency plasma thrusters and magnetic plasma nozzles.
\newblock {\em Reviews of Modern Plasma Physics}, 3:3, 2019.
\newblock \href {https://doi.org/10.1007/s41614-019-0024-2} {\path{doi:10.1007/s41614-019-0024-2}}.

\bibitem{BITT04}
J.A. Bittencourt.
\newblock {\em Fundamentals of plasma physics}.
\newblock Springer, Berlin, Germany, 2004.

\bibitem{domi18c}
A.~Dom{\'i}nguez-V{\'{a}}zquez, F.~Cichocki, M.~Merino, P.~Fajardo, and E.~Ahedo.
\newblock Axisymmetric plasma plume characterization with 2{D} and 3{D} particle codes.
\newblock {\em Plasma Sources Science and Technology}, 27(10):104009, 2018.
\newblock \href {https://doi.org/10.1088/1361-6595/aae702} {\path{doi:10.1088/1361-6595/aae702}}.

\bibitem{domi21a}
A.~Dom{\'i}nguez-V{\'{a}}zquez, F.~Cichocki, M.~Merino, P.~Fajardo, and E.~Ahedo.
\newblock On heavy particle-wall interaction in axisymmetric plasma discharges.
\newblock {\em Plasma Sources Science and Technology}, 30(8):085004, 8 2021.
\newblock \href {https://doi.org/10.1088/1361-6595/ac1715} {\path{doi:10.1088/1361-6595/ac1715}}.

\bibitem{nava18a}
J~Navarro-Cavall{\'e}, M~Wijnen, P~Fajardo, and E~Ahedo.
\newblock Experimental characterization of a 1 k{W} helicon plasma thruster.
\newblock {\em Vacuum}, 149:69--73, 2018.
\newblock \href {https://doi.org/10.1016/j.vacuum.2017.11.036} {\path{doi:10.1016/j.vacuum.2017.11.036}}.

\bibitem{thor09}
E.G. Thorsteinsson and G.T. Gudmundsson.
\newblock A global (volume averaged) model of a nitrogen discharge: I. steady state.
\newblock {\em Plasma Sources Science and Technology}, 18:045001, 2009.

\bibitem{gud13}
J.T. Gudmundsson, E.~Kawamura, and M.A. Lieberman.
\newblock A benchmark study of a capacitively coupled oxygen discharge of the oopd1 particle-in-cell {Monte Carlo} code.
\newblock {\em Plasma Sources Science and Technology}, 22:035011, 2013.

\bibitem{brown20}
Nathan~P. Brown and Mitchell L.~R. Walker.
\newblock Review of plasma-induced {H}all thruster erosion review of plasma-induced {H}all thruster erosion.
\newblock {\em Applied Sciences}, 10(11):3775, 2020.

\end{thebibliography}
\end{document}